\documentclass[prc,twocolumn,show pacs,preprintnumbers,amsmath,amssymb,altaffilletter,floatfix,superscriptaddress,,prc]{revtex4-1}
\usepackage{graphicx}
\usepackage{hyperref}
\usepackage{dcolumn}
\usepackage{bm}
\usepackage{float}
\usepackage{verbatim}
\usepackage[section]{placeins}
\usepackage[svgnames]{xcolor}
\usepackage[many]{tcolorbox}
\tcbuselibrary{listings}
\definecolor{ared}{rgb}{.647,.129,.149}

\definecolor{green}{rgb}{0.,0.5, 0.}

\definecolor{Dark}{gray}{.2}
\definecolor{MedDark}{gray}{.4}
\definecolor{Medium}{gray}{.6}
\definecolor{Light}{gray}{.8}

\definecolor{codegreen}{rgb}{0,0.6,0}
\definecolor{codered}{rgb}{0.6,0.1,0}
\definecolor{codegray}{rgb}{0.5,0.5,0.5}
\definecolor{codepurple}{rgb}{0.58,0,0.82}
\definecolor{backcolour}{rgb}{0.95,0.95,0.92}
\definecolor{lightgray}{gray}{0.95}
\definecolor{codeblue}{rgb}{0.117,0.403,0.713}

\newcounter{ipythcntr}
\renewcommand{\theipythcntr}{\texttt{[\arabic{ipythcntr}]}}
\newcommand{\ipin}[1][]{
    \stepcounter{ipythcntr}
    \hspace{-10pt}
    \color{codeblue}In  \theipythcntr}
\newcommand{\ipout}[1][\theipythcntr]{
    \hspace{-10pt}
    \color{codered}Out \theipythcntr}
\usepackage{color}
\definecolor{blue}{rgb}{0.,0., 1.}

\newcommand{\cspeed}{\mbox{$c$}}

\newcommand{\degree}{\mbox{$^\circ$}}

\newcommand{\gev}{\mbox{GeV}}
\newcommand{\gevtwo}{\mbox{GeV}^2}
\newcommand{\gevc}{\mbox{GeV/\cspeed}}

\newcommand{\gevctwo}{\mbox{(GeV/\cspeed)}^2}
\newcommand{\heep}{\mbox{$^1H(e,e^\prime)$}}
\newcommand{\deep}{\mbox{$^2H(e,e^\prime p)n$}}

\newcommand{\ceep}{\mbox{$^{12}C(e,e^\prime p)$}}

\newcommand{\qvec}{\mbox{$\vec q$}}

\newcommand{\ppi}{\mbox{$p_i$}}
\newcommand{\vppi}{\mbox{$\vec{p}_i$}}

\newcommand{\vppf}{\mbox{$\vec{p}_f$}}
\newcommand{\ppm}{\mbox{$p_{m}$}}
\newcommand{\vppm}{\mbox{$\vec{p}_{m}$}}

\newcommand{\xbj}{\mbox{$x_{B}$}}

\newcommand{\Qtwo}{\mbox{$Q^2$}}

\newcommand{\thetapq}{\mbox{$\theta_{pq}$}}
\newcommand{\thetanq}{\mbox{$\theta_{nq}$}}
\newcommand{\phip}{\mbox{$\phi_p$}}

\newcommand{\thetae}{\mbox{$\theta_e$}}

\newcommand{\dfsig}{\mbox{$\frac{\partial^5\sigma}{\partial\omega\partial\Omega_e\partial\Omega_{p}}$} }

\newcommand{\sigred}{\mbox{$\sigma_{red}$}}
\newcommand{\INFN} {INFN, Sezione Sanita and Istituto Superiore di Sanita, Laboratorio di Fisica, I-00161 Rome, Italy}
\newcommand{\ODU} { Old Dominion University, Norfolk, Virginia 23529} 
\newcommand{\Rutgers}{Rutgers, The State University of New Jersey, Piscataway, New Jersey 08854}
\newcommand{\JLAB}{Thomas Jefferson National Accelerator Facility,  Newport News, Virginia 23606}
\newcommand{\RENTEC}{present address: Renaissance Technologies LLC,  East Setauket, NewYork 11733}
\newcommand{\ANL}{ Argonne National Laboratory, Argonne, Illinois 60439}
\newcommand{\FIU}{Florida International University, University Park, Florida 33199}
\newcommand{\WM}{College of William and Mary, Williamsburg, Virginia 23187}
\newcommand{\CALSTATE}{California State University, Los Angeles, Los Angeles, California 90032}
\newcommand{\UMD}{University of Maryland, College Park, Maryland 20742}
\newcommand{\UVA}{University of Virginia, Charlottesville, Virginia 22901}
\newcommand{\UPS}{Universit\'e Paris-Saclay, CNRS/IN2P3, IJCLab, 91405 Orsay, France}
\newcommand{\LPSC}{Universit\'e Joseph Fourier, CNRS/IN2P3, INPG, Grenoble, France}

\begin{document}

\preprint{APS/123-QED}

\title{ $\deep$ Studies of Exclusive Deuteron Electro-Disintegration}

\author{W.U.~Boeglin} \affiliation{\FIU} 

\author{P.~Ambrozewicz} \altaffiliation{present address: \JLAB} \affiliation{\FIU} 
\author{K.~Aniol} \affiliation{\CALSTATE} 
\author{J.~Arrington}  \altaffiliation{present address: Lawrence Berkeley National Laboratory, Berkeley, California, 94720} \affiliation{\ANL}
\author{G.~Batigne} \affiliation{\LPSC}
\author{P.~Bosted} \affiliation{\JLAB} 
\author{A.~Camsonne} \affiliation{\JLAB} 
\author{L.~Coman} \altaffiliation{present address: Science Department, Florida Southwestern State College}\affiliation{\FIU} 
\author{G.~Chang} \affiliation{\UMD} 
\author{J.P.~Chen} \affiliation{\JLAB} 
\author{S.~Choi} \affiliation{Temple University, Philadelphia, Pennsylvania 19122} 
\author{A.~Deur} \affiliation{\JLAB} 
\author{M.~Epstein} \affiliation{\CALSTATE} 
\author{J.M.~Finn}\altaffiliation{deceased} \affiliation{\WM} 
\author{S.~Frullani} \altaffiliation{deceased}\affiliation{\INFN} 
\author{C.~Furget} \affiliation{\LPSC} 
\author{F.~Garibaldi} \affiliation{\INFN} 
\author{O.~Gayou} \affiliation{\JLAB} 
\author{R.~Gilman} \affiliation{\Rutgers}\affiliation{\JLAB}
\author{O.~Hansen} \affiliation{\JLAB} 
\author{D.~Hayes} \affiliation{\ODU}
\author{D.W.~Higinbotham} \affiliation{\JLAB} 
\author{W.~Hinton} \affiliation{\ODU} 
\author{C.E.~Hyde} \affiliation{\ODU} 
\author{H.~Ibrahim} \affiliation{Physics Department, Faculty of Science, Cairo University, Giza 12613,Egypt.}\affiliation{\ODU} 
\author{C.W.~de~Jager} \altaffiliation{deceased}\affiliation{\JLAB} 
\author{X.~Jiang} \affiliation{\Rutgers}
\author{M.~K.~Jones} \affiliation{\JLAB} 
\author{L.J.~Kaufman} \altaffiliation{present address: Indiana University, Bloomington, Indiana 47405} \affiliation{University of Massachusetts Amherst, Amherst, Massachusetts 01003}
\author{H.~Khanal} \altaffiliation{present address: Lockheed Martin, Rochester, NY}\affiliation{\FIU} 
\author{A.~Klein} \affiliation{Los Alamos National Laboratory, Los Alamos, New Mexico 87545} 
\author{S.~Kox} \affiliation{\LPSC} 
\author{L.~Kramer} \affiliation{\FIU} 
\author{G.~Kumbartzki} \affiliation{\Rutgers}
\author{J.M.~Laget} \affiliation{\JLAB} 
\author{J.~LeRose} \affiliation{\JLAB} 
\author{R.~Lindgren} \affiliation{\UVA} 
\author{D.J.~Margaziotis} \affiliation{\CALSTATE} 
\author{P.~Markowitz} \affiliation{\FIU} 
\author{K.~McCormick} \affiliation{Kent State University, Kent, Ohio 44242} 
\author{Z.~Meziani} \affiliation{Temple University, Philadelphia, Pennsylvania 19122} 
\author{R.~Michaels} \affiliation{\JLAB} 
\author{B.~Milbrath} \affiliation{\JLAB} 
\author{J.~Mitchell} \altaffiliation{\RENTEC} \affiliation{\JLAB}
\author{P.~Monaghan} \affiliation{Hampton University, Hampton, Viginia 23668} 
\author{M.~Moteabbed} \affiliation{\FIU} 
\author{P.~Moussiegt} \affiliation{\LPSC} 
\author{R.~Nasseripour} \affiliation{\FIU} 
\author{K.~Paschke} \affiliation{\UVA} 
\author{C.~Perdrisat} \affiliation{\WM} 
\author{E.~Piasetzky} \affiliation{Unviversity of Tel Aviv, Tel Aviv, Israel} 
\author{V.~Punjabi} \affiliation{Norfolk State University, Norfolk, Virginia 23504} 
\author{I.A.~Qattan} \affiliation{Khalifa University of Science, Technology and Research, Department of Physics, Sharjah, UAE}\affiliation{\ANL}
\author{G.~Qu\'em\'ener} \affiliation{\LPSC} 
\author{R.D.~Ransome} \affiliation{\Rutgers}
\author{B.~Raue} \affiliation{\FIU} 
\author{J.S.~R\'eal} \affiliation{\LPSC} 
\author{J.~Reinhold} \affiliation{\FIU} 
\author{B.~Reitz} \affiliation{\JLAB} 
\author{R.~Roch\'e} \affiliation{Ohio University, Athens, Ohio45701} 
\author{M.~Roedelbronn} \affiliation{University of Illinois, Urbana Champaign, Illinois 61820} 
\author{A.~Saha} \altaffiliation{deceased}\affiliation{\JLAB} 
\author{K.~Slifer} \affiliation{The University of New Hampshire, Durham, New Hampshire 03824} 
\author{P.~Solvignon} \altaffiliation{deceased}\affiliation{\ANL} 
\author{V.~Sulkosky}\altaffiliation{present address: Massachusetts Institute of Technology, Cambridge, Massachusetts 02139} \affiliation{\JLAB} 
\author{P.E.~Ulmer} \altaffiliation{\RENTEC}\affiliation{\ODU}
\author{E.~Voutier}  \altaffiliation{\UPS}\affiliation{\LPSC} 
\author{L.B.~Weinstein} \affiliation{\ODU} 
\author{B.~Wojtsekhowski} \affiliation{\JLAB} 
\author{M.~Zeier} \affiliation{\UVA} 

\collaboration{For the Hall A Collaboration}

\date{\today}

\begin{abstract}

  The $\deep$ cross section was measured at momentum transfers $\Qtwo = $ 0.8, 2.1 and 3.5 $\gevctwo$
  covering a wide range of proton kinematics at each $\Qtwo$ setting that made it possible to study
  this reaction as a function of missing momentum as well as a function of
  the neutron laboratory recoil angle $\thetanq$.  Missing momentum
  distributions were determined for fixed values of $\thetanq$ up to missing momenta of 0.65 $\gevc$. For the two larger
  momentum transfer settings, the characteristics of the experimental momentum distributions 
  confirm the theoretical prediction that final state interactions (FSI) contribute maximally 
  around a $\thetanq \sim 70\degree$, while for $\thetanq < 45\degree$ FSI are significantly reduced. 
  The data at reduced FSI settings were best reproduced by calculations using the CD-Bonn potential wave functions. 
  
\end{abstract}

\maketitle
%
\section{Introduction}\label{L_introduction}
The understanding of the short-range structure of the deuteron is of fundamental importance for the advancement of our understanding of nuclear matter at short distances. To probe the short-range properties of the deuteron, one has to investigate configurations where the two nucleons come very close together and are strongly overlapping. 
The basic problem is to what extent these configurations can be described simply in terms of two nucleons with high initial relative momenta. The ultimate quantity to be investigated in this case is the high momentum component of the deuteron wave function.
Traditionally three classes of reactions are used to study the high momentum part of the deuteron wave function: elastic scattering, inclusive and exclusive electro-disintegration reactions.

Elastic electron-deuteron scattering probes the integrated characteristics of the wave function via the deuteron form-factors. In this case the only free kinematic parameter is the 4-momentum transfer $-Q^2$.
As the analysis of  experimental data~\cite{Alexa} showed, at presently available energies, it is practically impossible to discriminate between different theoretical approaches~\cite{mgwvo, GilGro02} used to calculate the deuteron elastic form-factor $A(Q^2)$. 
One needs additional constraints on the deuteron wave function at short distances that can be provided by inelastic electron scattering reactions.

The onset of short range dynamics has been clearly established in dedicated quasi-elastic inclusive electron scattering experiments on the deuteron and other nuclear targets~\cite{Rock82,Rock92,Arrington_1998ps}.
In this case the cross section depends on a bound nucleon momentum distribution integrated over its transverse momentum component. For the case of a deuteron target the extracted quantity is the y-scaling function $F(y)$ \cite{Arrington_1998ps,Fomin:2011ng}, 
 where $y$ is related to the 
longitudinal component of the nucleon momentum. However, it is currently impossible to precisely quantify 
inelastic contributions (which increase with $Q^2$) and effects of final state interactions (FSI) (which increase with $\xbj$ for $\xbj > 1$, where $\xbj$ is the Bjorken scaling variable) rendering the relation of extracted $F(y)$ function to 
the underlying high momentum distribution of the deuteron more uncertain.   

The most direct way of probing high momentum nucleons in the deuteron is to investigate the quasi-elastic (QE) electro-disintegration of the deuteron via the $\deep$ reaction in knock-out kinematics and at high missing momenta (the momentum of the recoiling neutron) $\vppm = \qvec - \vppf$, where $\vppf$ is the momentum of the observed proton and $qvec$ is the transferred momentum. The knock-out kinematics of this reaction ensures that the outgoing proton carries almost all of the transferred momentum $\vppf \sim \qvec$. 
Within the Plane Wave Impulse Approximation (PWIA) $-\vppm$ corresponds to the pre-existing momentum $\vppi$ of the bound nucleon. Thus the strategy in these studies is to probe the cross section at $\ppm$ values as large as possible. 
Depending on the selected kinematics, this scattering process can be overwhelmed by final state interactions  where the knock-out proton interacts with the recoiling neutron, charge interchange FSI, or by processes where the virtual photon couples to the exchanged meson (MEC) or where  the nucleon is excited to an intermediate $\Delta$ state (IC). 

The dominance of FSI, MEC and IC has seriously affected previous experiments at $\Qtwo < 1$ $\gevctwo$ ~\cite{edsacley,bl98,boe08,Ulm02,Rva05,Boeglin:2015cha,Arrington:2011xs} leading to  the overall conclusion that these experiments do not provide an effective access to the high momentum components of the deuteron  wave function.


Theoretical studies of high energy exclusive electrodisintegration reactions\cite{FSS96,treeview} indicate that 
the condition $\Qtwo \geq 1$ $\gevctwo$ is necessary in order to enhance contributions of reaction mechanisms which probe the short-range structure of the deuteron and to suppress competing long range processes for the following reasons:  
(i) the MEC contribution will be suppressed  by an additional factor of  $(1+Q^2/\Lambda)^{-4}$  with $\Lambda = 0.8-1$ $\gevctwo$ as compared to the QE contribution~\cite{treeview,hnm}; 
(ii) the  large $Q^2$ limit will allow one to probe the wave function in the $\xbj>1$ region which is far from the inelastic threshold, thereby suppressing IC contributions; 
(iii) final state interactions of the outgoing nucleon will follow the eikonal dynamics with a strong angular anisotropy dominating mainly  in transverse  direction with respect to $\qvec$.  This situation  generated a multitude of theoretical studies  of  the $\deep$  reaction in the high $Q^2$ regime~\cite{Sabine1,FMGSS95,FSS96,Jesch2001,Ciofi08,treeview,JW08,JW09,Kap05,La05,jw10,noredepn}.
The PWIA predictions of these studies (see particularly Refs.~\cite{FSS96,treeview,noredepn,La05,JW08})  were largely different due to the use of different deuteron wave functions and the treatment of  off-shell effects. However,  they all agreed in the suppression of FSI effects at  the level of 10-20\% in the angular range of $30 \le \theta_{nq} \le 45^0$ for the spectator neutron production relative to the momentum transfer.

In the current paper we present the results of the analysis of the full data set on the $\deep$ reaction covering a wide kinematical range with $Q^2 = 0.8, 2.1$ and $3.5$ $\gevctwo$.  In addition to verifying the FSI suppression effects and attempting to extract the ``genuine" deuteron momentum distribution up to large missing momenta, our analysis also makes it possible to establish the onset of the eikonal regime which is essential for 
setting the stage for exploring of the deuteron at very short distances.
The wide range in momentum transfer, missing momenta and neutron recoil angles allows us to access the high momentum components of the deuteron
momentum distribution and probe the validity of current models of the reaction dynamics.  

Part of the kinematic region covered in the current experiment overlaps with a $\deep$ experiment performed using CLAS at Jefferson
Lab~\cite{Kim3} which concluded that FSI and IC are dominating the momentum distribution except for $\ppm < 0.1$ $\gevc$ or $\thetanq > 110\degree$.   
However, to obtain reasonable statistical precision, the data were integrated over the full $\thetanq$ range for the momentum distributions  and over a large $\ppm$-range for the angular distributions in contrast to the data presented below. These integrations did not permit the observation of the above mentioned kinematic window where FSI are suppressed and thereby providing access to the high momentum component of the deuteron momentum distribution. 

In sections \ref{L_experiment} and ~\ref{L_analysis} we present details of the experiment and its analysis, respectively.
In section~\ref{L_theory} describes the theories that are used in the data analysis and compared to the new experimental data.
In section~\ref{L_results} an overview of the experimental results is presented and  contains a detailed comparison and discussion of the new results with theoretical calculations. Section~\ref{L_summary} presents a summary and conclusion.

\section{Experiment}\label{L_experiment}

The experiment was carried out using the two high-resolution spectrometers (HRS) in Hall A at Jefferson Lab. 
The HRS on the left side relative to the direction on the incoming beam, detected the scattered electrons and the  HRS on the right side the ejected protons. The deuterium target consisted of a 15~cm long cylinder which was filled with liquid deuterium.  
An identical target cell filled with liquid hydrogen was used for calibration purposes and to determine the coincidence efficiency.  
In addition to the cryogenic targets, solid carbon targets, one consisting of several carbon foils were used for spectrometer optics optimization and solid aluminum targets simulating empty cryotarget  cells (so called dummy targets) were installed to evaluate contributions of the target walls to the recorded events. 

The electron beam was rastered over an area of $2\times2$ mm$^2$ and the cryogenic liquids were continuously circulated in order to minimize density variations due to boiling. In spite of these measures a sizable average reduction of the effective target thickness  as a function of beam intensity was observed which is described in detail in Section~\ref{L_analysis}.


The spectrometer detection systems in the two arms were very similar:
vertical drift chambers (VDC) were used for tracking and two scintillation counter planes (S1/S2) following the VDCs provided timing and trigger signals. In addition, the electron arm was equipped with a gas {\v C}erenkov detector for e$^-$/$\pi^-$ discrimination. We found that the gas {\v C}erenkov detector was sufficient for the $\pi^-$ rejection in this experiment. In general, we found that $\pi$ background was not a concern in these measurements.
A detailed description of the spectrometer and the target systems can
be found in reference~\cite{halla}.
\begin{figure*}[htp]
\includegraphics[width=0.8\textwidth,clip=true]{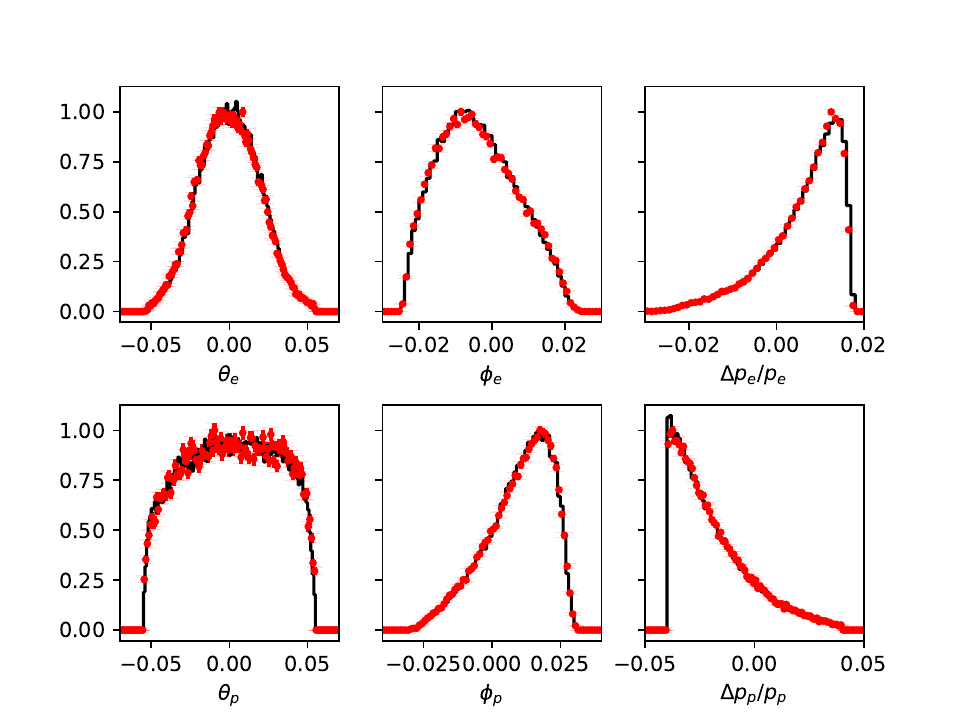}
\caption{ (Color online) Comparison between the simulated yield using SIMC (solid line) and the experimental data (red data points with error bars) for $\Qtwo = 2.1$ $\gevctwo$ and $\ppm = 0.1$ $\gevc$. The simulated data have been scaled to give the same integrated yield as the data.}
\label{fig:SIMC_comp}
\end{figure*} 

$\deep$ cross sections were measured at four-momentum transfers of $\Qtwo=0.8,\ 2.1$ and 3.5 $\gevctwo$ at beam energies of $E_{inc} = 2.843,\ 4.703$ and 5.008 $\gev$, respectively. 
For each fixed momentum transfer the kinematic settings were centered at missing momentum values of $\ppm= 0.0,\ 0.1,\ 0.2,\ 0.4$ and $ 0.5 $ $\gevc$, while the angle $\thetanq$ of the recoiling neutron with respect to the momentum transfer in the laboratory frame $\qvec$, was varied. $\thetanq$ is also referred to hereafter as the recoil angle.   Varying the recoil angle while keeping $\ppm$ and $\Qtwo$ constant, required the energy transfer, the electron scattering angle, the proton final momentum and the  proton spectrometer angle to be adjusted accordingly for each value of $\thetanq$.  
For a fixed missing momentum there is a one-to-one correspondence between the recoil angle $\thetanq$ and $\xbj$. For the various kinematic settings, $\xbj$ varied therefore between  0.52 and 1.81, for $\Qtwo=0.8$ and $2.1$ $\gevctwo$,  and between 0.83 and 1.53, for $\Qtwo=3.5$ $\gevctwo$. Note that large $\xbj$ values correspond to small recoil angles and vice versa.

Overall a total of 65 electron/proton spectrometer settings were combined to extract recoil angle distributions at fixed missing momentum as well as missing momentum distributions at fixed recoil angle.


\clearpage
\section{Analysis}\label{L_analysis}
The momentum acceptance used for both
spectrometers was limited by software cuts to $\Delta p /{p_0}$
=$\pm$4~$\%$, where $p_0$ is the central momentum of the spectrometer and $\Delta p = p - {p_0}$.

The solid angle of each spectrometer was defined by software cuts at
the entrance of the first quadrupole magnet.
In addition a second,  global cut on the multi-dimensional acceptance of each spectrometer
was applied by means of R-functions~\cite{Rfunc}.

As both spectrometers can reconstruct the reaction vertex, we also placed a cut on the difference in the location of the reaction vertex along the beam. This difference has typically a full width half maximum (FWHM) of about 0.8 - 1.2 cm. 
An additional cut was applied on the location of the reaction vertex such that the effective target length was reduced to 10~cm instead of the full 15~cm provided by the target cell. This cut eliminated contributions from the cell walls and also avoided the edges of the long-target acceptance of the spectrometers.

\begin{figure}[H]
\includegraphics[width=0.46\textwidth,clip=true]{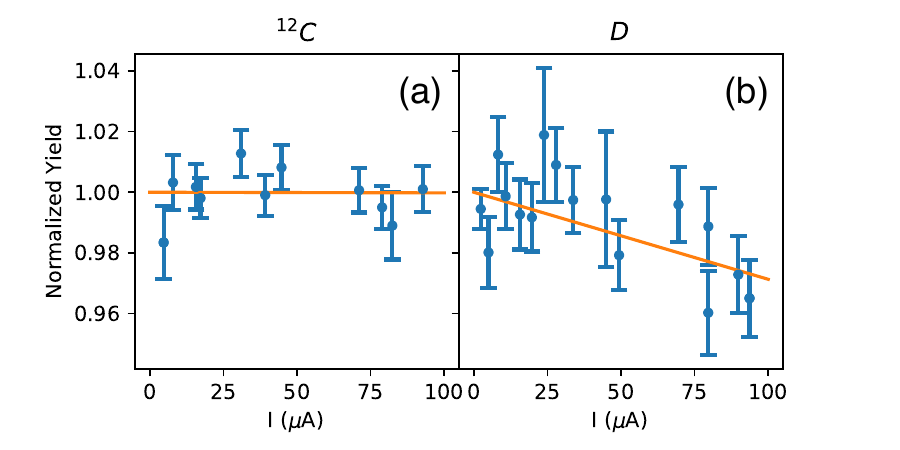}
\caption{ (Color online) Relative variation of the target thickness as a function of beam current: (a) $\ceep$, (b)$\deep$ rates are corrected for trigger and tracking efficiencies.}
\label{fig:lumi}
\end{figure}

 Fig.~\ref{fig:lumi} shows the effect of deuterium target thickness variation as a function of beam current compared to the $\ceep$ rates for which no variation with beam current is expected when corrected for dead time and detector inefficiencies. The lines are linear fits, for $\ceep$ at 100~$\mu$A we find a target thickness of $100.0 \pm 0.5\%$ relative to 0 $\mu$A and for $\deep$ we find a relative target thickness of $97.1 \pm 0.6\%$.    For the hydrogen target the effective target thickness dropped to  $70.0\pm 0.8\% $ at the same maximal current. The particle yields were corrected on a run-by-run basis for this effective target thickness variation dependent on the average current for the respective run. The error associated with this correction was added in quadrature to the statistical error of the data.
The experimental yields were further corrected for detector inefficiencies on a run-by-run basis.

For an overall coincidence efficiency measurement we determined elastic cross sections from $\heep$ coincidence data and compared them to the published values from the fit of Table I in ref.~\cite{Arr04}. The cross sections were measured at beam currents below 20$\mu$A to minimize  effective target thickness variations. The coincidence efficiencies were found to be $97.\pm 5.5\%$ for the $\Qtwo=0.8$ and $2.1$ $\gevctwo$ data set and $96.\pm 4.4\%$  for the $\Qtwo=3.5$ $\gevctwo$ data set. The larger error bar in the lower $\Qtwo$ data set is due to larger fluctuations in the determined $\heep$ cross sections as a function of beam current. The uncertainties quoted above also include the uncertainty in the published cross section values. These are scale uncertainties that apply to the overall set of measurements at each $\Qtwo$.

The systematic error due to uncertainties in the measured and reconstructed kinematic variables were calculated for each data bin.  An additional combined systematic error of 4.5\% was determined for beam
charge measurements, detector efficiencies, absolute target thickness and cut variations. The two systematic errors were combined quadratically with the statistical error.

For the $\Qtwo=0.8$ and $2.1$ $\gevctwo$ data sets the phase space acceptance as well as the radiative corrections were calculated using a version of the Monte-Carlo code SIMC~\cite{SIMC, Ent01}. For the $\Qtwo=3.5$ $\gevctwo$ data set, 
the phase space acceptance was calculated using the Hall A Monte-Carlo
code MCEEP~\cite{MCEEP} while the extracted cross sections were radiatively corrected using  SIMC. MCEEP was used for historical reasons and the phase space acceptance calculated with MCEEP and SIMC were found to be identical within their respective statistical uncertainties. For all data sets the simulated particle yield was estimated using a theoretical calculation by J.M. Laget~\cite{La05} that included final state interactions and reproduced the experimental yield quite well. It was found that the variation in the radiative corrections as a function of the $\deep$ reaction model used (e.g. using PWIA or include FSI) was less than 1\%. The Monte-Carlo calculations describe the shapes of the spectrometer acceptances very well as is illustrated in Fig.~\ref{fig:SIMC_comp} where the experimental distributions of reconstructed transport variables at the target are compared to the calculated ones. The red data points correspond to the experimental data and the solid lines represent the calculated distributions from SIMC. The simulated distributions have been scaled such that the respective integrated yields are identical.

\clearpage
\section{Theoretical Calculations}\label{L_theory}

Theoretical calculations play a twofold role in this analysis: (i) the new experimental data are ultimately used to test state of the art theoretical models of the $\deep$ reaction at high momentum transfer and (ii) they are also needed in the analysis of the data and in the interpretation of the new results. During data analysis, calculations are needed to determine corrections to the raw data (e.g. radiative correction and acceptance correction calculations) and simpler models can be used where the error introduced by this simplification are small and can be estimated.  

The simplest model of the $\deep$ reaction is the Plane Wave Impulse Approximation (PWIA) where the incoming as well as the scattered electrons are described as plane waves and the ejected proton is described by a plane wave as well. This means that a bound proton absorbs a virtual photon and leaves the nucleus without further interaction.  In this case the cross section can be factorized and written as:
\begin{equation}
\dfsig = \kappa \, \sigma_{ep} \, f_{rec} \, \rho(p_m)
\label{eq:pwia}
\end{equation}
where $\kappa$ is a kinematic factor, $\sigma_{ep}$ is the off-shell electron-proton cross section for a moving nucleon, $f_{rec}$ is the recoil factor and $\rho(\ppm)$ is the momentum distribution corresponding to the square of the  deuteron wave function in momentum space.  In this case the initial momentum $\vppi$ of the bound nucleon is directly related to the momentum of the recoiling system $\vppm$ by $\vppi = -\vppm$ which in turn can be determined from 3-momentum conservation via $\vppm = \qvec - \vppf$, where $\vppf$ is the final momentum of the ejected proton.

As mentioned in the introduction when FSI are present, the ejected nucleon re-interacts with the residual system and the direct correlation between the initial momentum and the recoil momentum is lost. A similar process also occurs due to MEC and IC. Current state-of the art models of the $\deep$ reaction include final state interactions but mostly neglect MEC and IC as they are expected to be considerably suppressed at high $\Qtwo$. 
In the following sections detailed comparisons between the experimental data and theoretical calculations are presented. These calculations are using wave functions derived from several NN-potential models. The main features of these potentials are summarized in the following.

The Paris potential model~\cite{Paris_Potential}  is based on one-pion-exchange (OPE), correlated and uncorrelated two-pion-exchange and $\omega$-exchange contributions. It includes a purely phenomenological part for short distances $r \leq 0.8$~fm where the radial dependence is constant and energy dependent. The available NN scattering data were then fitted with a relatively small number of parameters. As this form was not suitable for many-body calculations a purely phenomenological parameterization was created and fitted to the available experimental data. This purely mathematical form can be used in many-body calculations and was used to calculate the deuteron wave function labeled Paris below.

The CD-Bonn potential~\cite{CD_Bonn_Potential}  is also a one boson exchange model based on an effective, field-theoretical approach where all known mesons below the nucleon mass are allowed to contribute. This model uses Feynman amplitudes of meson exchange in their original form without any local approximations. It was shown (Fig.2 of ~\cite{CD_Bonn_Potential}) that the half off-shell $^3$S$_1$-$^3$D$_1$ potential is quite sensitive to this approximation for off-shell momenta above 0.3-0.4~$\gevc$. This also results in a different D-state probability compared to the other models and can lead to differences in the momentum distribution for nucleon momenta above 0.3 $\gevc$. The CD-Bonn potential includes boson parameters that are partial wave dependent.

The WJC2 potential~\cite{GROSS2007176}  is a OBE model calculated within the framework of the Covariant Spectator Theory. The scattering amplitude is obtained from the solution of a covariant integral equation (Gross Equation) derived from field theory.  The WJC2 potential was able  to fit the available NN scattering data with very high quality using only 15 fit parameters. Part of its success has been attributed by the authors to automatically including relativistic structures that are difficult to identify and very difficult to add to non-relativistic models without adding additional fit parameters.

The AV18 potential~\cite{AV18_Potential} by the Argonne group is a phenomenological potential that represents a compromise between a restricted OBE model and a very flexible OBE model where parameters are allowed to vary for each partial wave. The AV18 has a local operator structure but is not restricted by a OBE form for short distances. It has been optimized to produce a virtually perfect fit to the Nijmegen NN scattering data base.  

Wave functions derived from the potentials described above have been used in the following calculations of the $\deep$ cross sections:

\begin{enumerate}
	\item M.~Sargsian's  $\deep$ cross section calculations ~\cite{FSS96,treeview,noredepn} are using wave functions derived from the CD-Bonn potential~\cite{CD_Bonn_Potential} (referred to as MS PWIA CDB or MS FSI CDB below ) as well as wave functions calculated with the V18 potential~\cite{AV18_Potential} (referred to as MS PWIA V18 or MS FSI V18). The transition amplitude is calculated within the framework of the virtual nucleon approximation  and derived in covariant form from effective Feynman diagrams. This approximation considers only the $pn$ component of the deuteron  and is valid for large momentum transfers $\Qtwo > 1$ $\gevctwo$ and for kinetic energies of the recoil nucleon $T_N <$~0.5~$\gev$.  It is also assumed that the virtual nucleon operator contributes negligibly to the scattering amplitude, which is satisfied for recoil momenta $\ppm \leq 0.7$ $\gevc$.  Final state interactions have been described as forward elastic re-scatterings of the struck nucleon using the generalized eikonal approximation (GEA) which, in contrast to the conventional Glauber approximation, includes the Fermi motion of the bound nucleons. 
		
	\item Cross sections from J-M.~Laget's model~\cite{La05}, referred to as JML PWIA or JML FSI, are using Paris potential~\cite{Paris_Potential} wave functions. In this calculation the transition amplitude is expressed as a series of dominant diagrams, computed in momentum space in the Lab frame. The propagators as well as the kinematics involved have been kept fully relativistic. The elementary vertex operators have been adjusted to reproduce experimental results in the relevant channels. This method has been discussed for the $\deep$ reaction in refs.~\cite{JML_CJP_1984_1046,JML_NPA_1978_265}. 
This model also implicitly includes the Fermi motion of the bound nucleons when FSI are calculated. While the virtual nucleon approximation is valid for large $\Qtwo$ values this model includes a continuous transition from a low-energy description of FSI in terms of partial waves to a high-energy description with a mostly absorptive amplitude similar to the one used in M.~Sargsian's calculation.  
    As mentioned previously JML was also used to calculate radiative corrections and bin center corrections. 
    	
    \item Calculations by W.~Ford,  S.~Jeschonnek and J.W.~van Orden ~\cite{PhysRevC.90.064006} determine cross sections using a fully relativistic formalism that is based on a Bethe Salpeter equation approximation together with wave functions from the WJC2~\cite{GROSS2007176} and the CD-Bonn~\cite{CD_Bonn_Potential} potentials. Final state interactions are included by expressing the FSI re-scattering amplitude with a Fermi invariant parameterization that include all possible spin dependencies of the $np$ scattering matrix. 
    The necessary invariant functions have been determined either in terms of the SAID partial wave analysis data (valid for laboratory kinetic energies up to 1.3 GeV) or using a Regge based model fit to all available $NN$ scattering data  for energies 5.4 $\leq  s \leq$ 4000~$\gevtwo$.  In kinematic regions where both $NN$ data sets are available good agreement between these two methods was found in most observables. This model therefore allows one to calculate $\deep$ cross section over a very wide range of kinematic settings. Cross sections from these calculations will be referred to as FJO PWIA WJC2 or FJO FSI WJC2 and FJO PWIA CDB or FJO FSI CDB depending on the wave functions used.

\end{enumerate}.


\clearpage
\section{Results}\label{L_results}

The large kinematic coverage of the combined kinematic settings made it possible to study two different aspects of the $\deep$ reaction  
for each $\Qtwo$ setting. We determined angular cross section distributions for fixed missing momenta of $\ppm= 0.0,\ 0.1, \, 0.2, \,0.4,$ and  0.5 $\gevc$ as a function of the angle of the recoiling neutron with respect to the momentum transfer $\thetanq$  in the Lab frame (the recoil angle) and we  measured $\deep$ cross sections as a function of missing momentum while the recoil angle $\thetanq$ was kept constant. A list of all 65 spectrometer settings is provided in appendix~\ref{app:A}.
Angular distributions are especially suitable to investigate contributions of final state interactions which are expected to be  very anisotropic at large momentum transfers (see ref.~\cite{FSS96,treeview,noredepn}). Missing momentum distributions on the other hand are closely related to the deuteron wave function in momentum space. As a first overview of the data sets, in section~\ref{L_R_overview} the entire set of cross section data is presented graphically; section~\ref{L_R_angdist}, focuses on the analysis of the angular distributions for the five, fixed values of missing momentum listed above and in section~\ref{L_R_momdist} the missing momentum dependence of the cross section for a range of fixed recoil angles will be presented and discussed.

\subsection{Cross Section Overview}\label{L_R_overview}
\begin{figure}[htp]
\includegraphics[width=0.4\textwidth,clip=true]{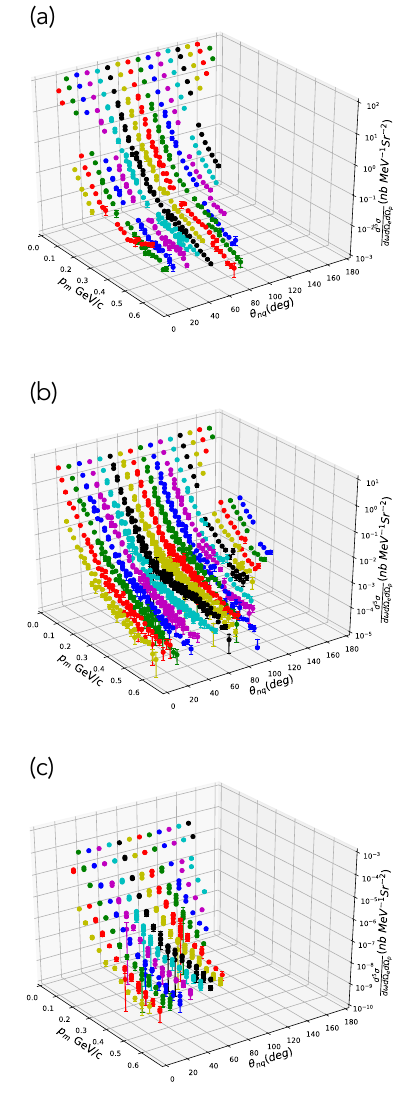}
\caption{ (Color online) Experimental cross sections as a function of missing momentum $\ppm$ and recoil angle $\thetanq$ for $\Qtwo = 0.1$ (a), $\Qtwo = 2.1$ (b) and $\Qtwo = 3.5$ $\gevctwo$ (c). The various colors represent different values of $\thetanq$.}
\label{fig:sigma}
\end{figure}

To provide an overview of the complete data set, experimental cross sections as a function of $\thetanq$ and missing momentum are shown in Fig.~\ref{fig:sigma}. The colored groups of cross section data indicate data sets at constant recoil angles where only the missing momentum varies. For each momentum transfer setting the cross sections vary by about five orders of magnitude. For several recoil angle values the missing momentum range extends beyond 0.5 $\gevc$. At momentum transfers 2.1 and 3.5 $\gevctwo$ (see Fig.~\ref{fig:sigma} (b) and (c)) and at a recoil angle around 75$\degree$ (black data points) the cross sections are typically considerably larger than for smaller as well as larger recoil angles, this behavior is due to the dominance of final state interaction in this kinematical region and will be discussed in more detail below. At the larger momentum transfers one can observe clusters of data points for the same missing momentum ($\ppm$) value which is a consequence of different spectrometer settings contributing to the same $\ppm\thetanq$ bin and is further discussed in the following sections.  Numerical cross section values together with their kinematic settings are provided in the supplemental materials~\cite{supp} and a description of the data provided is given in appendix~\ref{app:B}..   

\subsection{Angular Distributions}\label{L_R_angdist}
The kinematic settings used in this experiment were selected in such a way that their respective central values correspond to missing momentum values of $\ppm= 0.0,\ 0.1,\ 0.2,\ 0.4$ and $ 0.5 $ $\gevc$ and for each, fixed $\ppm$-value a set of $\xbj$ values with their corresponding $\thetanq$-values was assigned. 

\begin{figure}[htp]
\includegraphics[width=0.4\textwidth,clip=true]{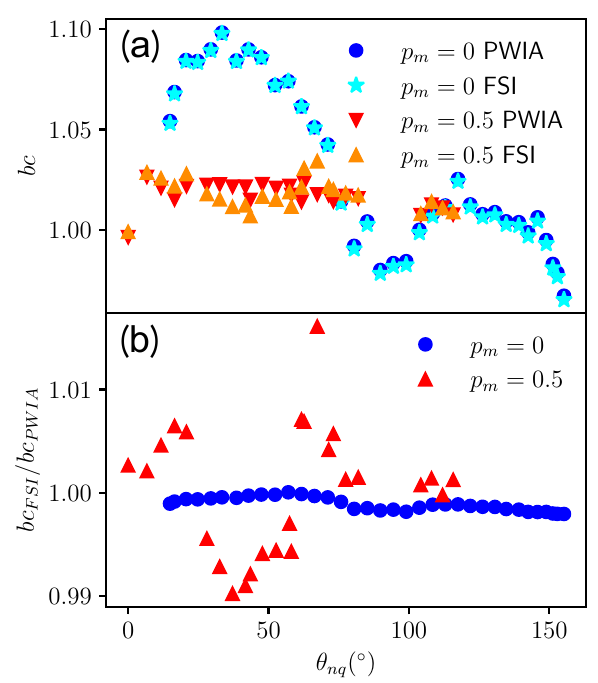}
\caption{ (Color online) (a) Bin correction factors ($f_{bc}$) for the angular distributions for $\ppm = 0$ GeV/c (blue circles: FSI, violet stars: PWIA) and for $\ppm = 0.5$ GeV/c (orange upward triangles: FSI, red downward triangles: PWIA). (b) The ratio of calculated bin correction factors including FSI ($f_{bc_{FSI}}$) to bin correction factors using PWIA only ($f_{bc_{PWIA}}$). Blue circles $\ppm = 0$ GeV/c, red triangles $\ppm = 0.5$ GeV/c.}
\label{fig:bc_calc}
\end{figure}

The finite momentum and angular acceptances of the two HRS spectrometers result in a range of kinematic values ($\ppm, \thetanq, \thetae, \phip$, where $\thetae$ is the electron scattering angle and $\phip$ is the angle between the electron scattering plane and the reaction plane) that were probed at each spectrometer setting and are referred to as the phase space acceptance of the two spectrometers. 

Multiple spectrometer settings with frequently overlapping acceptances contributed to the full angular distribution for a given missing momentum. The condition of constant $\Qtwo$ and constant $\ppm$ was maintained for the central part of each spectrometer acceptance only. 
Within the phase space acceptance of a specific spectrometer setting the kinematic variables varied slightly around their central values as a function of the angle $\thetanq$. This led to variations of the experimental cross section as a function of $\thetanq$ that were independent of reaction mechanism effects for small variations. We therefore determined the cross sections for a small but finite range of missing momentum values ($\Delta\ppm = \pm 0.02$ $\gevc$) centered at the nominal central setting (also referred as the missing momentum bin) and for a small recoil angle range ($\Delta \thetanq = \pm 2.5\degree$) centered at $\thetanq$ (the recoil angle bin).  
All events whose $\ppm$ and $\thetanq$ values were in these ranges contributed to the measured cross section for this kinematic bin. Cross section variations within these bins are mostly due to small changes of the electron scattering angle and the missing momentum and can lead to differences between the averaged cross section and the cross section for the averaged kinematic setting for this bin. 
Using a realistic model of the $\deep$ reaction (in this case JML PWIA and JML FSI) the averaged cross section $\overline{\sigma_i}$ and the the averaged kinematics $\overline{kin_i}$ for each bin $i$ where calculated using the Monte-Carlo code SIMC described previously. The so-called bin-correction factor $f_{bc}$ is then determined as shown in eq. \eqref{eq:fbc} below, where $\sigma_{\overline{kin_i}}$ is the cross section calculated at the averaged kinematics of bin $i$. Multiplying the experimental cross section with $f_{bc}$ results in a realistic estimate of the experimental cross section at the averaged kinematics of this bin $i$ and makes it possible to compare the measured cross sections to  different theoretical models without the need of including the models in a Monte-Carlo simulation. The model dependence introduced that way
\begin{equation}
f_{bc} =  \frac{\sigma_{\overline{kin_i}}}{\overline{\sigma_i}}
\label{eq:fbc}
\end{equation}
 is shown in Fig.~\ref{fig:bc_calc} using  a set of  representative bin correction factors. For the smallest missing momentum settings and at the edge of the phase space acceptance $f_{bc}$ can be up to 10 - 15\%, while for large missing momenta they are typically smaller than 5\%, (Fig.~\ref{fig:bc_calc} panel (a)).  Fig.~\ref{fig:bc_calc} panel (b) shows the model dependence of the bin correction factors. This has been estimated by comparing their values calculated using the JML FSI model to those from the JML PWIA model. Overall the model dependence of the correction factors is below 1\% for small $\ppm$ values and increases up to 1.5\% for large $\ppm$ values where FSI dominate. This model dependence has been included in the total systematic error estimate.

To study the angular distribution (as a function of $\thetanq$) of the experimental cross sections, we divided them by the corresponding PWIA cross sections (see eq.~\eqref{eq:pwia}) where $\kappa$ is a kinematic factor,  $f_{rec}$ is the recoil factor,  $\rho(\ppm)$ refers to the momentum distribution calculated with the Paris potential and $\sigma_{cc1}$ is the de~Forest CC1 off-shell cross section~\cite{defor83} calculated using the form factor parameterization from Table I of Ref.~\cite{Arr04}. This procedure has the advantage that those parts of large cross section variations  which are well understood within the PWIA framework are removed and deviations from the simple PWIA model are enhanced. 
The resulting ratios $R(\thetanq) = \sigma_{exp}/\sigma_{pwia}$ were then averaged for common $\thetanq$ bins of overlapping spectrometer settings.


%
\begin{figure}[htp]
\includegraphics[width=0.50\textwidth,clip=true]{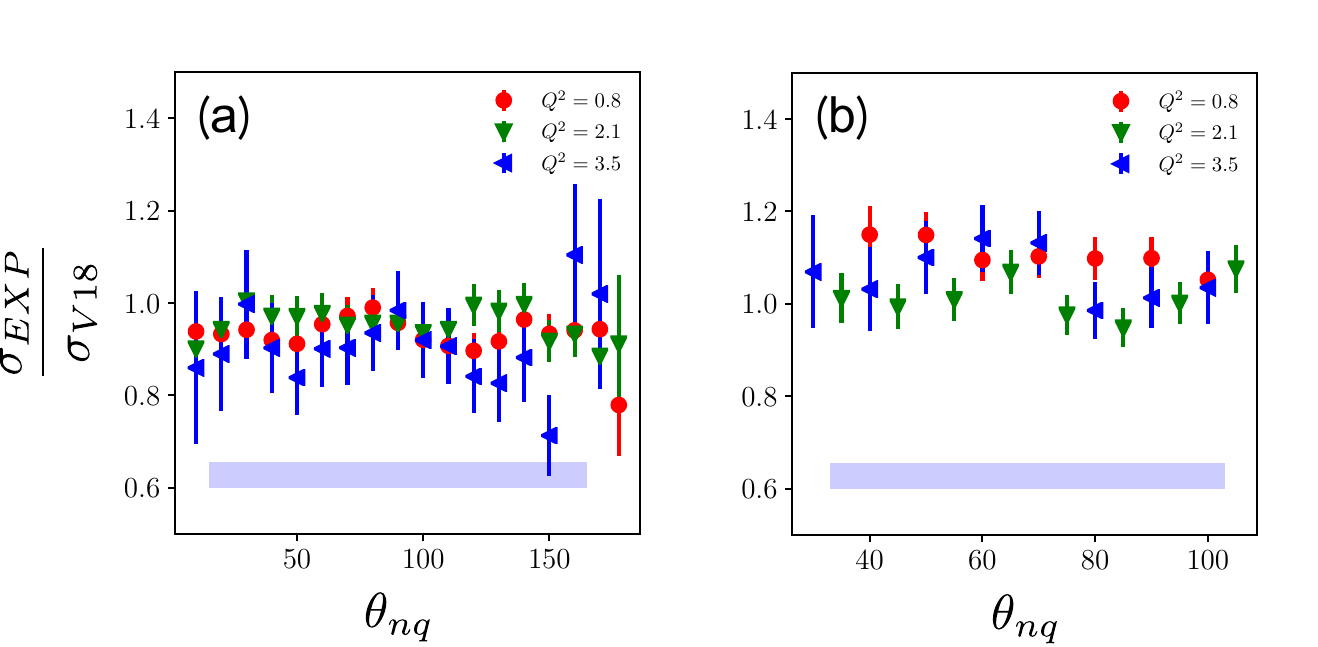}
\caption{ (Color online) Ratio of experimental cross sections to calculated cross section using the MS FSI V18 model (described in Sec.~\ref{L_theory}). (a) For $\ppm = 0.0$ $\gevc$ and (b) for $\ppm = 0.1$ $\gevc$. The red circles are for $\Qtwo = 0.8$, the green triangles (pointing down) for $\Qtwo = 2.1$ and the blue triangles (pointing left) are for $\Qtwo = 3.5$ $\gevctwo$. The horizontal bar indicates the overall normalization error for the data sets. }
\label{fig:sargs_q2dep}
\end{figure} 

If FSI were completely absent, then a PWIA model would describe the data and this ratio would be independent of the recoil angle and proportional to the ratio of the true momentum distribution to the one used in the calculation. 
Low missing momentum regions can be used to check the consistency of cross sections measured at different beam energies and  momentum transfers provided that FSI contributions are small. These ratios should be identical for all data low $\ppm$ data sets and show only small variations if FSI calculations are taken into account.
Calculating FSI contributions is inherently model dependent but within a model one can check their dependence on $E_i$ and $\Qtwo$. As an example Fig.~\ref{fig:sargs_q2dep} shows $R = \sigma_{exp}/\sigma_{V18}$ from the MS FSI V18 calculation by M.~Sargsian for missing momentum settings of $\ppm = 0.0$ and $0.1$ $\gevc$ and for the three measured momentum transfers. There is no clear $\Qtwo$ dependence of this ratio and the experimental data appear to be consistent within their total and systematic errors.

\begin{figure*}[htp]
\includegraphics[width=1.0\textwidth,clip=true]{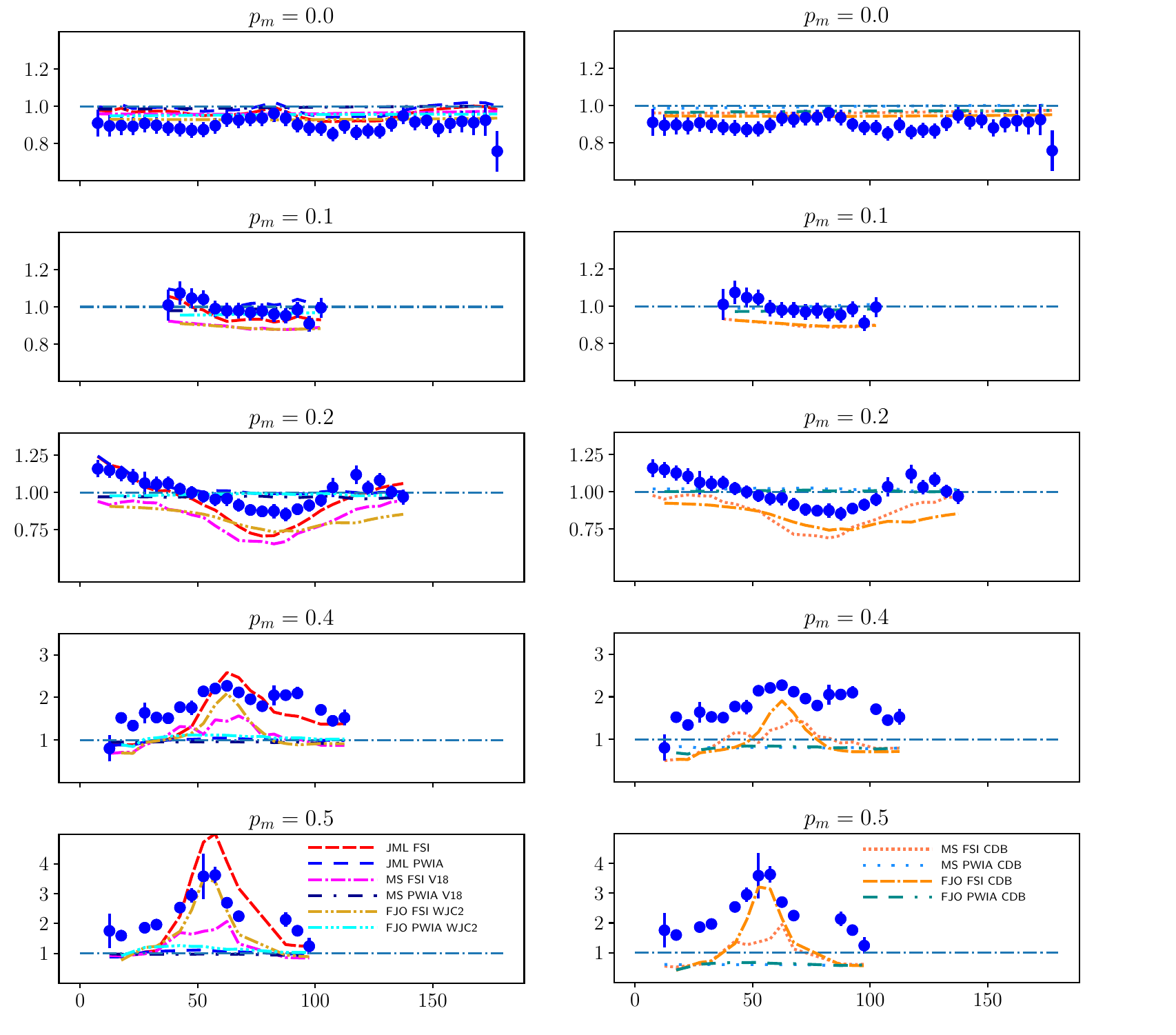}
\caption{ (Color online) Ratios of experimental  to theoretical PWIA cross sections  for $\Qtwo = 0.8$ $\gevctwo$ as a function of  $\thetanq$ for various central missing momentum settings. On the left are calculations using the V18 and WJC2 potentials and on the right the cross sections have been calculated using the CD-Bonn potential. See section~\ref{L_theory} for a desciption of these calculations.}
\label{fig:r_pwia08}
\end{figure*}

\begin{figure*}[htp]
\includegraphics[width=1.0\textwidth,clip=true]{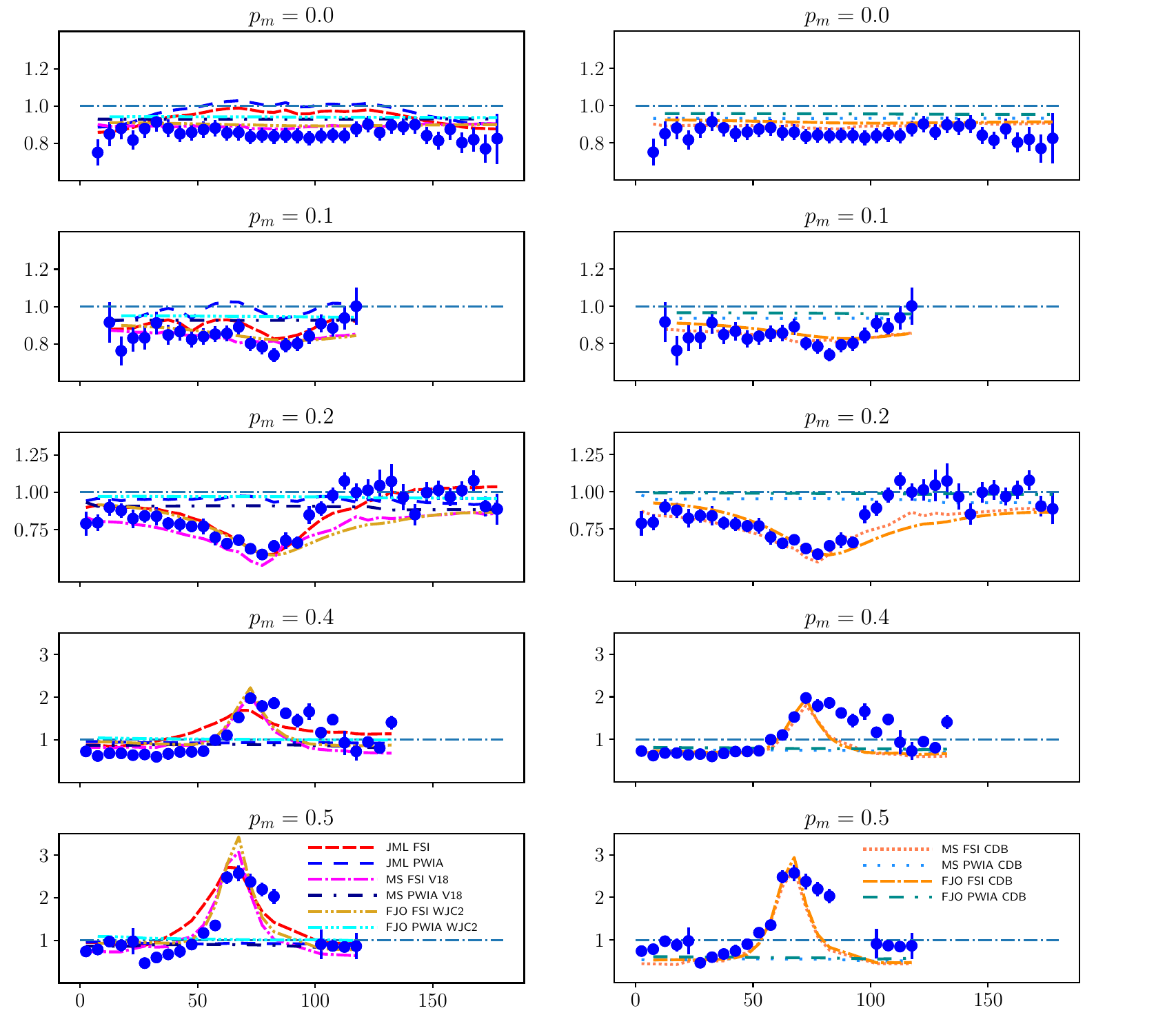}
\caption{ (Color online) Ratios of experimental  to theoretical PWIA cross sections for $\Qtwo = 2.1$ $\gevctwo$ as a function of  $\thetanq$ for various central missing momentum settings. On the left are calculations using the V18 and WJC2 potentials and on the right the cross sections have been calculated using the CD-Bonn potential. See section~\ref{L_theory} for a description of these calculations.}
\label{fig:r_pwia21}
\end{figure*}

\begin{figure*}[htp]
\includegraphics[width=1.0\textwidth,clip=true]{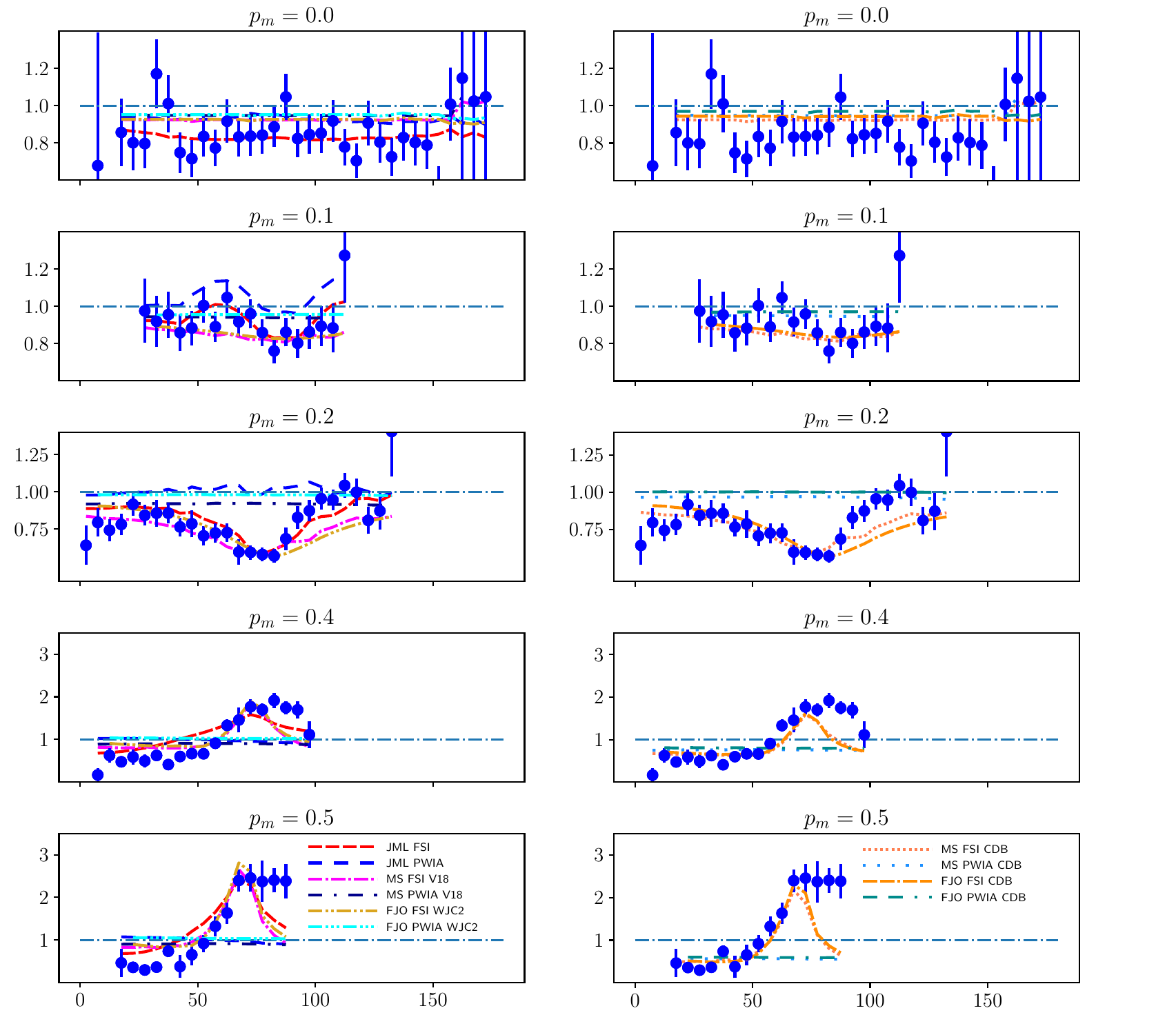}
\caption{ (Color online) Ratios of experimental  to theoretical PWIA cross sections  for $\Qtwo = 3.5$ $\gevctwo$ as a function of  $\thetanq$ for various central missing momentum settings. On the left are calculations using the V18 and WJC2 potentials and on the right the cross sections have been calculated using the CD-Bonn potential. See section~\ref{L_theory} for a description of these calculations.}
\label{fig:r_pwia35}
\end{figure*}

The final angular distributions for all five missing momentum settings at the three momentum transfers settings are shown together with theoretical calculations in Figs.~\ref{fig:r_pwia08}, \ref{fig:r_pwia21} and~\ref{fig:r_pwia35}, respectively. The calculations are with and without final state interactions to highlight the effect of FSI contributions. We also separated calculations using the V18 and WJC2 wave deuteron functions from those based on the CD-Bonn wave function for clarity. All theoretical cross sections were calculated at the averaged kinematic setting for each bin and then processed identically to the experimental cross sections i.e.   
the ratio $R = \sigma_i/\sigma_{PWIA}$ where $\sigma_i$ represents the experimental or a calculated cross section and $\sigma_{PWIA}$ is the PWIA cross section calculated using the momentum distribution calculated with the Paris potential and the de Forest CC1 off-shell cross section~\cite{defor83}.

For all three $\Qtwo$, at small missing momenta centered around $\ppm = 0.0$ and $0.1$ $\gevc$ there is indeed little structure, indicating that the experimental cross sections follow relatively closely a PWIA description and FSI and other effects are of the order of 10\% and angle independent. 
For the smallest missing momentum settings $\ppm = 0$  and $0.1$ $\gevc$ and at all momentum transfers, the experimental data as well as the calculations show virtually no $\thetanq$ dependence. In general the PWIA calculations for the JML, MS and FJO calculations using wave functions from the Paris, V18 and CD-Bonn potentials are in agreement with each other within typically 5\%. They overestimate the experimental reduced cross sections typically by 5 - 15\% depending on the recoil angle. Contributions due to FSI are found to be at the level of about 5-15\%, reducing the cross section (compared to the PWIA calculation) and improving the agreement with experiment. The MS CDB and the FJO CDB calculations are generally very close ($\lesssim 5$\%) to each other with and without FSI for all missing momentum values. One can therefore summarize that at these small missing momentum regions FSI are small and recoil angle independent. If FSI are included all calculations are in agreement with the experimental data at the 5-10\% level. For $\Qtwo$=2.1~$\gevctwo$  we find that the JML FSI agrees best with the data. For $\Qtwo = 2.1$ all FSI calculations by MS and FJO agree while JML FSI results are systematically somewhat over predicting,   and at $\Qtwo = 3.5$ $\gevctwo$ JML FSI agrees slightly better than all MS and FJO calculations including FSI. 
For missing momenta centered around $\ppm = 0.2 \pm 0.02$ $\gevc$, the distributions for $\Qtwo = 2.1$ and 3.5 $\gevctwo$ show a clear reduction of $R$ for $\thetanq$ around 75\degree ($\xbj$ $\sim$1), while for the smallest momentum transfer setting a considerably smaller variation can be observed with only an indication of a dip at around 80\degree. 
The reduction in cross section observed for the higher momentum transfers is an indication that the kinematic regime has been reached where FSI can be successfully described using the eikonal approximation. This FSI description~\cite{treeview,hnm,Sabine1,FMGSS95,FSS96,Jesch2001,Ciofi08}  predicts a strong angular dependence of FSI, dominating near transverse recoil directions of $\thetanq \approx 75\degree$.

For the $\Qtwo = 0.8$ $\gevctwo$ data set none of the calculations agree with the experimental data very well. For $\ppm = 0.2$ $\gevc$ they all predict a larger dip in $R$ around $\thetanq \approx 80\degree$ compared to what is observed.  For larger missing momenta, $\ppm = 0.4$ and 0.5 $\gevc$,  JML with FSI predicts a much more peaked structure for $\thetanq \approx 80\degree$ with areas of small FSI for recoil angles smaller than 40\degree which is not observed. MS-CDB as well as MS-V18 with FSI under-predict $R$ by a factor of two. The failure of reproducing the general trend of $R(\thetanq)$ at this momentum transfer is an indication that at this kinematic setting the eikonal regime has not been reached and the underlying assumptions for this calculation have not been satisfied. On the other hand this seems to be a kinematic regime where the particle energies involved are too large for a non-relativistic calculation to be fully valid as well. 
 
At the larger momentum transfer settings the agreement with all calculations including FSI improve dramatically for $\thetanq \leq 80\degree$. For JML FSI there is very good agreement for $\ppm = 0.2$ $\gevc$ over the full range while the other calculations systematically lie below the experimental values for  $\thetanq \gtrsim 80\degree$ by about 20\%.

 MS-CDB is overall in excellent agreement with the data for $\ppm = 0.4$ and 0.5 $\gevc$ for $\thetanq \leq 80\degree$. The exception is at $\ppm = 0.5$ $\gevc$ for $\Qtwo = 2.1$ $\gevctwo$ and recoil angles smaller than 25\degree where the experimental cross section increases considerably. As discussed in section~\ref{L_R_momdist} below, this trend can also be observed in the momentum distributions for this range of recoil angles and missing momenta above 0.4 $\gevc$ and seems to be decreasing with increasing momentum transfer. At these small recoil angles the center of mass (CM) energy of the n-p system is minimal which could introduce FSI which are not described by the eikonal approximation.

This is  further reinforced when looking at the angular distributions for missing momenta $\ppm = 0.4$ $\gevc$ and $\ppm = 0.5$ $\gevc$ $R$ shows a peak at around $75\degree$ with a maximal value of $\sim1.6$ and $\sim2.5$, respectively. 
The dependence of $R$ on $\thetanq$ at large $\Qtwo$ is also in agreement with the measurement of Ref.~\cite{Kim3} for missing momenta $0.4 \leq \ppm \leq 0.6$ $\gevc$.

The angular distributions for $\Qtwo = 0.8$ $\gevctwo$ in contrast are very broad and do not show a strong angular dependence. In addition, the absolute value of the ratio for missing momenta $\ppm > 0.4$ $\gevc$ is quite big indicating substantial FSI contributions over the entire angular distribution.

Thus our data confirm that the strong angular dependencies for $\Qtwo = 2.1$ and 3.5 $\gevctwo$ indicate that the struck proton is energetic enough that its interaction with the recoiling, slow neutron is in the eikonal or geometric regime where it is well known that the slow particle recoils almost in a perpendicular direction with respect to the fast particle. The much smoother angular dependence at $\Qtwo = 0.8$ $\gevctwo$ can be interpreted that at these energies the eikonal regime has not been completely reached and substantial FSI are present for all recoil angles and missing momenta of 0.4 $\gevc$ and above.

It is also interesting to observe that for the higher momentum transfers, the values of $R$ for the various missing momenta seem to have the same value for $\thetanq \approx 30 - 45\degree$ indicating a region where FSI are independent of missing momenta and, according to all calculations, small. This makes these kinematic regions ideal for probing the momentum distribution of the proton in the ground state wave function.

The observed reduction of FSI is a consequence of the eikonal approximation and the behavior of the high energy nucleon-nucleon scattering amplitude. In the on-shell part of the re-scattering amplitude $A_R$, the pole in the  knockout nucleon propagator (see eq. 29 and 30 in~\cite{noredepn}) leads to an explicit factor $i$. At the same time the high energy $NN$ on-shell scattering amplitude is predominantly imaginary (see eq. 31 in~\cite{noredepn})  leading to a final on-shell amplitude that has a negative sign relative to the Born/PWIA amplitude. Consequently the total scattering cross section will contain an interference term with a negative sign:
$$
(A_I + A_R)^2 = |A_I|^2 - 2 |A_I A_R| + |A_R|^2
$$
where $A_I$ is the direct (PWIA) amplitude. At small missing momenta $A_R$ is much smaller than $A_I$ and FSI are generally small. With increasing missing momenta $A_R$ increases in such a way that while $A_R^2 \ll A_I^2$ the interference term $2 A_I A_R > |A_R|^2$ leading to the overall reduction in cross section observed at around $\ppm \approx 0.2$ $\gevc$. At larger missing momenta $A_R$ can start to dominate leading to large contributions to the cross sections depending on the recoil angle $\thetanq$. Since $A_R$ has a strong angular dependence, due to the dominance of the re-scattering at transverse directions, there exist angular ranges where the re-scattering term $|A_R|^2$ and the interference term $2 |A_I A_R|$ cancel each other such that the only remaining contribution comes from the direct term $A_I^2$ for which PWIA based calculation become valid.  

Common to all calculations is that the measured values of $R$ for $\thetanq \geq 80\degree$ are systematically larger than all calculations. This can likely  be attributed to intermediate $\Delta$-isobar excitation as for these data points the kinematic setting is such that it would favor $\Delta$ excitation on a nucleon at rest instead of hitting a high momentum proton.

Data files containing the numerical results of the data shown in Figs.~\ref{fig:r_pwia08},\ref{fig:r_pwia21} and \ref{fig:r_pwia35} together with detailed kinematic variables can be found in the supplemental information~\cite{supp}.

The problem of the description of the scattering process from a bound nucleon, generally 
referred to as off-shell effects, depends on the theoretical framework of the description of 
the scattering process itself. In the virtual nucleon approximation~\cite{noredepn} which has been used 
for the comparison with the data, off-shell effects are taken into account by assigning to the interacting nucleon $i$, an initial energy $E^{off}_i$=$m_D$-$E_s$, where the energy $E_s$=$\sqrt{m_N^2 + p_i^2}$ of the spectator nucleon is considered on-shell. In such an approach the propagator of 
the interacting nucleon is split to on- and -off shell parts in which the latter is proportional to a virtuality parameter, $\tau$= $({\sqrt{m_N^2 + p_i^2} - E^{off}_i )/ (2m_N)}$, which is treated 
as an off-shell parameter. The off-shell part of the propagator results in an off-shell component of electromagnetic current. The uncertainty of this off-shell component is reduced by requiring an overall current 
conservation 
 in which case the unknown longitudinal component of the current, $J_{N,z}$ is expressed through the $J_{N,0}$ component. The latter is constrained by the overall charge of the deuteron which is an observable and makes it possible to perform a self-consistent calculation of off-shell effects which are proportional to the ${\tau m_N}/\sqrt{Q^2}$-parameter. 
As such for a given virtuality $\tau$ of the nucleon, off-shell effects diminish with an increase of $Q^2$. Since off-shell effects are larger in the longitudinal component of the electromagnetic current, one expects the larger effects in kinematic settings with a large longitudinal component of the interacting nucleon momentum.  Thus, off-shell effects are expected to be relatively larger in parallel and antiparallel kinematics.

\begin{figure*}[htp]
\includegraphics[width=0.8\textwidth,clip=true]{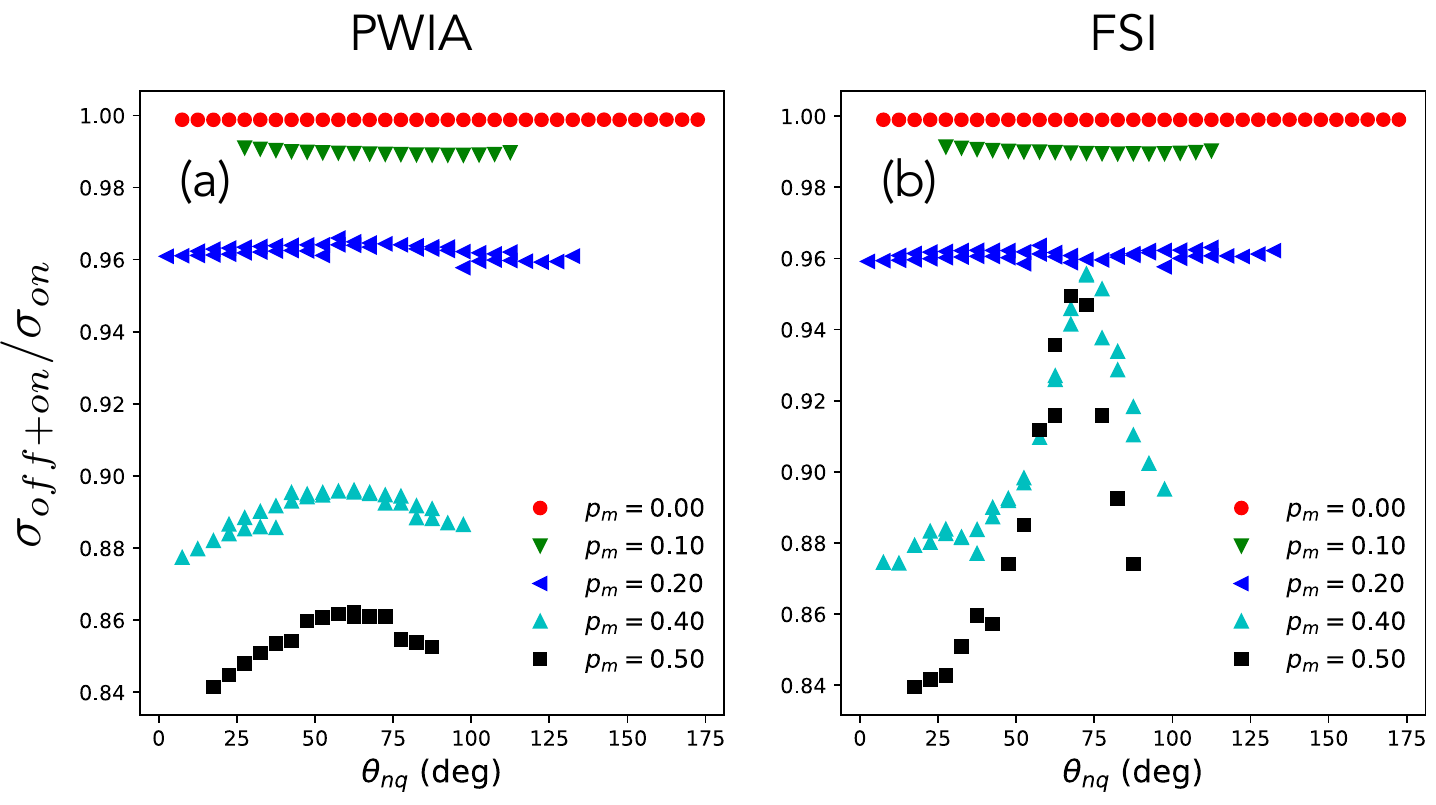}
\caption{ (Color online) Ratio between M. Sargsian calculations with ($\sigma_{off + on}$) and without ($\sigma_{on}$) off-shell effects for $\Qtwo= 3.5$ $\gevctwo$ and $\ppm = $ 0, 0.1, 0.2, 0.4 and 0.5 $\gevc$ (a) MS PWIA CDB, (b) MS FSI CDB}
\label{fig:r_os35}
\end{figure*}

We will use M. Sargsian's MS PWIA CDB and MS FSI CDB calculations to demonstrate the effect of the off-shell approximation described above. This calculation provides cross sections where the off-shell part of the interaction is neglected. For the comparison we used the large momentum transfer data  kinematics ($\Qtwo = 3.5$ $\gevctwo$) where the eikonal regime has been fully reached. Fig.~\ref{fig:r_os35} shows the cross section ratio between a calculation including off-shell effects ($\sigma_{off+on}$) and one containing only on-shell contributions ($\sigma_{on}$). Panel (a) is for the PWIA calculation and panel (b) shows the results when including FSI. One can see that in both calculations with increasing missing momentum the off-shell effects increase from about 1\% at $\ppm = 0.1$ $\gevc$ to about 16\% at $\ppm = 0.5 $ $\gevc$. In the PWIA calculation this ratio varies as a function of the neutron recoil angle reflecting the dependence of off-shell contributions on the parallel component of the nucleon momentum mentioned above. Including FSI and at $\thetanq$ angles $< 25\degree$ the behavior is similar to the PWIA one. At larger angles however the effect of off-shell contributions varies rapidly as a function of $\thetanq$, decreasing when $\thetanq$ approaches $\sim 75\degree$ and increasing again for larger angles. At $\thetanq \sim 75\degree$ the off-shell effect is approximately the same as for a missing momentum of $\ppm = 0.2$ $\gevc$ independent of the actual recoil momenta. The reason is that for $\thetanq \sim 75\degree$ the re-scattering term dominates the cross section and originates from a much lower nucleon momentum state of $\ppi \sim 0.2$ $\gevc$.  At this kinematics the re-scattering process transfers about 0.3 $\gevc$ of momentum between the struck and spectator nucleons, resulting in a final missing momentum of 0.4 - 0.5 $\gevc$.

\clearpage

\subsection{Momentum Distributions}\label{L_R_momdist}

\begin{figure*}[htp]
\includegraphics[width=0.95\textwidth,clip=true]{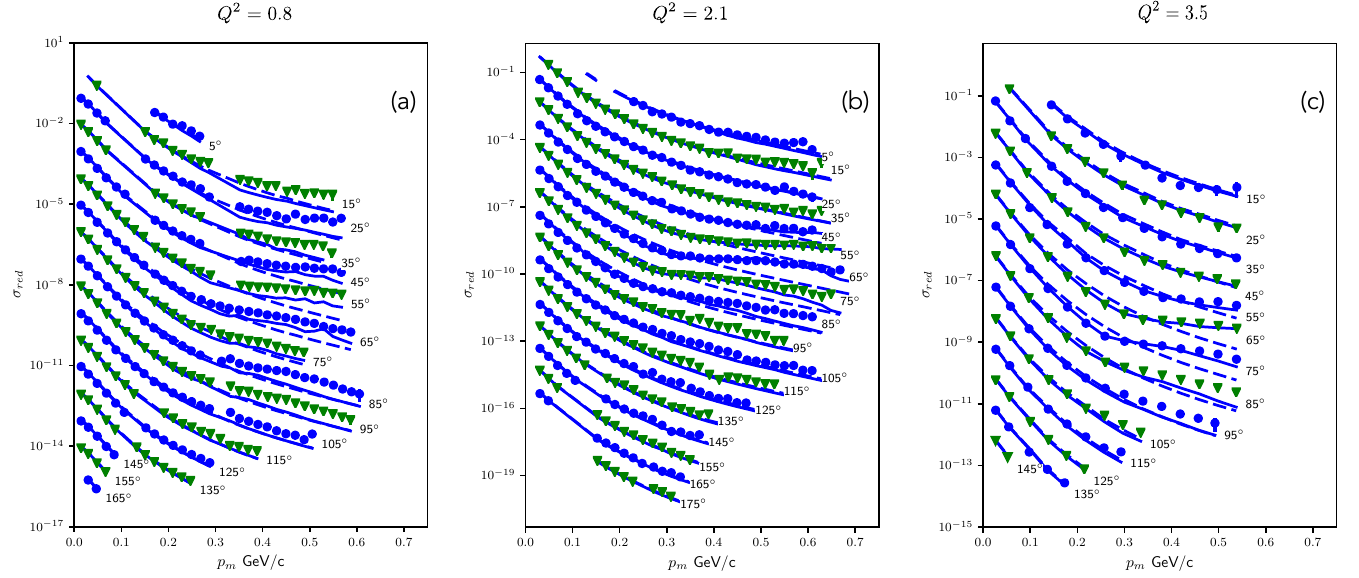}
\caption{ (Color online) The reduced cross section in fm$^3$
  $\sigma_{red}(\ppm)$ as a function
  of missing momentum $\ppm$ is shown in panels a, b and c for
  $\Qtwo = 0.8, 2.1$ and $3.5$ $\gevctwo$, respectively, at a fixed $\thetanq$ bins (between $5\degree$ to $175\degree$) with
  a bin width of $\pm 5\degree$. The experimental data are alternatively represented as filled blue circles and dark cyan downward pointing triangles. Calculations: MS PWIA CDB dashed
  lines  (blue) , MS FSI CDB solid (blue). The experimental data points include error bars which are most of the time smaller than the symbols used}
\label{fig:pm_dist}
\end{figure*}

Even though the focus of the original experiment was the determination of angular distributions at  selected missing momenta, the coincidence acceptance of the two HRS made it possible to determine $\deep$ cross sections over a wide range of missing momenta and recoil angles. We therefore binned the data in $\thetanq$ and $\ppm$ and determined the cross section for each  $\thetanq, \ppm$ bin.  We calculated the averaged kinematic variables and bin correction factors for each $\thetanq, \ppm$ bin in the same manner as described above for the angular distributions. We then determined reduced cross sections according to eq. \eqref{eq:sig_red}, 

\begin{equation}
\sigred = \dfsig \big/ ( \kappa \, \sigma_{ep} \, f_{rec})
\label{eq:sig_red}
\end{equation}
where $\kappa$ is a kinematic factor,  $f_{rec}$ is the recoil factor and $\sigma_{ep}$ is the de~Forest CC1 off-shell cross section~\cite{defor83}. If there were no FSI $\sigred$ would correspond closely to the momentum distribution. Reduced cross sections have the advantage that large variations of the cross section e.g. due to variations of the electron kinematics within the considered bin have been removed and FSI and off-shell effects do not vary very much within each kinematic bin. Since the same $\thetanq, \ppm$ bin can contain contributions from different spectrometer settings,  a weighted average of the experimental reduced cross sections was calculated from all spectrometer settings contribution to it. 
Fig.~\ref{fig:pm_dist} presents an overview of the experimental reduced cross sections as a function of missing momentum for sets of constant $\thetanq$ compared to the MS PWIA CDB (dashed) and MS FSI CDB calculations (solid). Scale factors (starting at $f = 1$ for the smallest recoil angle and decreasing subsequently by a factor of 10) were applied to each momentum distribution at successive recoil angles to separate them vertically and make their overall change in  shape clear. Several trends are visible: (i) at $\Qtwo = 0.8$ $\gevctwo$ FSI contribute significantly for $\ppm > 0.25$ $\gevc$ for all angles and no regions with reduced FSI are discernible within the spectrometer acceptance; (ii) in contrast the $\Qtwo = 2.1$ $\gevctwo$ data are quite well reproduced by the calculation without FSI for  $\ppm < 0.5$ $\gevc$ and for $\thetanq < 45\degree$. FSI are also small for $\thetanq> 95\degree$ but the data seem to be systematically above the calculation; (iii) at $\thetanq = 65, 75\degree$ and as a function of missing momentum one first observes a reduction in cross section compared to PWIA which reaches its minimum value at around $\ppm = 0.25$ $\gevc$ and then a continuous increase  for missing momenta of $\ppm > 0.4$ $\gevc$ and above which is further confirmation that the eikonal regime has been reached.   
The same overall pattern is also  observed for the $\Qtwo = 3.5$ $\gevctwo$ data set. The $\ppm$ bin width for this data set is 0.04 $\gevc$ in contrast to the 0.02 $\gevc$ for the lower $\Qtwo$ data sets. This change in bin width was to prevent statistical uncertainties from dominating. Again for angles $\thetanq > 95\degree$ FSI are predicted to be small but experimental cross sections are significantly larger than the calculated ones, indicating the dominance of other processes that have not been included in the calculations (e.g. virtual $\Delta$ excitation). 
\clearpage
Calculating the ratio between the experimental and the theoretical cross sections for each kinematic bin and spectrometer setting we present a more detailed comparison of the experimental cross sections to the theoretical ones below. 

\begin{figure*}[!p]
\includegraphics[width=0.95\textwidth,clip=true]{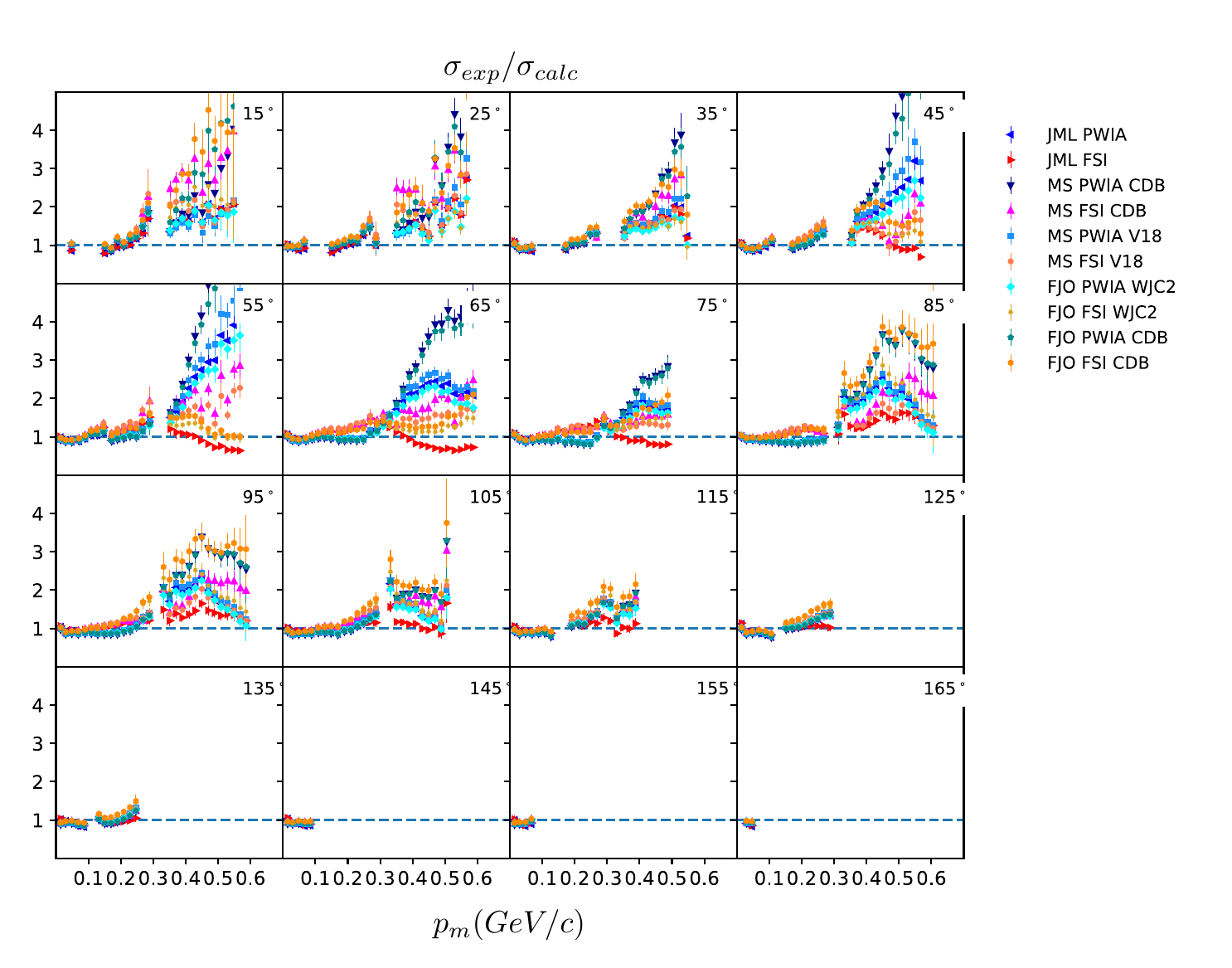}
\caption{ (Color online) Ratios of the experimental  to calculated cross sections  for a momentum transfer of $\Qtwo = 0.8$ $\gevctwo$. 
Cool, bluish colors and black represent PWIA calculations and warm, reddish colors are calculations including FSI. Specifically: left pointing triangles (dark blue) PWIA JML, right pointing triangles (red) JML FSI, down pointing triangles (black) MS PWIA CDB, upwards pointing triangles (pink) MS FSI CDB FSI, squares (blue) MS PWIA V18, circles (dark orange) MS FSI V18, diamond (cyan) FJO PWIA WJC2, star (ochre) FJO FSI WJC2, pentagon (dark cyan) FJO PWIA CDB and hexagon (orange) FJO FSI CDB. For a discussion of the various calculations see section ~\ref{L_theory}. The angles indicate the neutron recoil angle with respect to $\qvec$ for each data set.}
\label{fig:pm_ratio_q1}
\end{figure*}

For missing momentum bins with contributions from multiple spectrometer settings, we determined the weighted mean value of this ratio and its uncertainty. As in the angular distributions the averaged kinematic variables have been used in the theoretical calculation for each bin. For $\Qtwo = 0.8$ $\gevctwo$  the results are shown in Fig.~\ref{fig:pm_ratio_q1}. For missing momenta below 0.2 $\gevc$ the different calculations (including or excluding FSI) agree with the data within 10-20\% but deviate increasingly with increasing $\ppm$ for all the remaining angles. Especially CD-Bonn based calculations MS FSI CDB and FJO FSI CDB show large deviations for missing momenta above 0.3 $\gevc$ and $\thetanq$$\leq$35$\degree$. This overall discrepancy between calculation and experiment over a wide angular and missing momentum range reinforces the argument that at this momentum transfer setting the eikonal regime has not been reached and the calculations are not expected to represent the data well. In general the JML FSI calculations agree best with the data for recoil angles $\thetanq = 75^\circ$ and $\thetanq$$\geq$105$^\circ$ while there remain significant differences at other angles. One can conclude that this momentum transfer corresponds to a transition region where low energy calculations are not reliable anymore and high energy approximations are not yet applicable.


\begin{figure*}[htp]
\includegraphics[width=0.95\textwidth,clip=true]{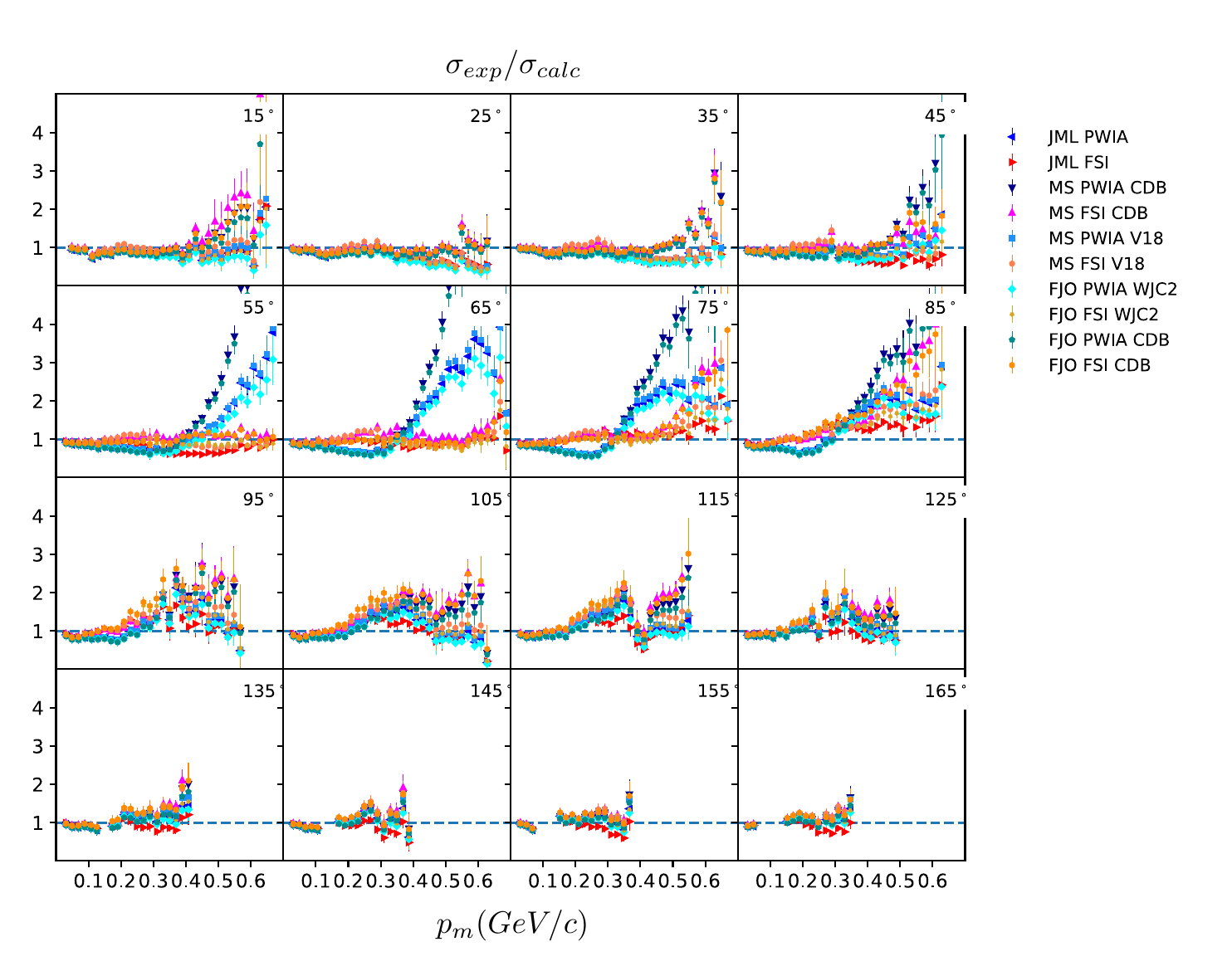}
\caption{ (Color online) Ratios of the experimental to calculated cross sections for a momentum transfer of $\Qtwo = 2.1$ $\gevctwo$. See Fig.~\ref{fig:pm_ratio_q1} for an explanation of the legends. The angles indicate the neutron recoil angle with respect to $\qvec$ for each data set.}
\label{fig:pm_ratio_q2}
\end{figure*}

In contrast to the low momentum transfer setting,  the $\Qtwo = 2.1$ $\gevctwo$ data set (Fig.~\ref{fig:pm_ratio_q2}) shows a much improved agreement between experiment and all calculations up to about $\ppm = 0.3$ $\gevc$ and $\thetanq \leq 45^\circ$. 
Common to all calculations is that FSI are predicted to be small for missing momenta up to 0.4 - 0.5 $\gevc$ and $\thetanq \leq 45^\circ$. Both MS FSI V18 and MS CDB as well as the FJO FSI WJC2 and FJO FSI CDB calculations indicate a sensitivity to the wave function for missing momenta between 0.3 and 0.5 $\gevc$. Especially at $\thetanq = 25\degree$ and $35\degree$ the data seem to agree better with the CD-Bonn calculation compared to the V18, WJC2 or Paris based calculations. For very small recoil angles ($<25\degree$) and missing momenta around 0.5 $\gevc$ and above, MS FSI CDB and FJO FSI CDB start to deviate significantly from the data with increasing missing momentum as the measured cross sections are larger than predicted. In addition, the ratio to the MS FSI V18 and the FJO FSI WJC2 calculations is larger at $15\degree$ compared to $25\degree$ for this missing momentum range. While CD-Bonn based calculations deviate from the data at $15\degree$ and $\ppm \geq 0.5$ $\gevc$ the V18 and WJC2 based calculations seem to agree better at this recoil angle setting. 
Corresponding behavior has been previously observed in the angular distributions for \ppm = 0.4 and \ppm = 0.5 $\gevc$ (Fig.~\ref{fig:r_pwia21}) and has been attributed to the small CM energy in the final $n$-$p$ system. Large missing missing momenta and very small recoil angles correspond to low relative $n$-$p$ energies. At  $\Qtwo = 2.1$ $\gevctwo$ these $n$-$p$ energies are at the edge or even outside of the kinematic region where the eikonal approximation is expected to be valid.  With increasing missing momentum at these small recoil angles the system moves further away from the eikonal regime resulting in larger deviations between experiment and calculation. 

At a recoil angle of $\thetanq = 45\degree$ FSI appear to be important for missing momenta above 0.5~$\gevc$. FSI contributions continue to grow dramatically for larger recoil angles and also affect the lower missing momentum regions. As observed in the angular distributions FSI lead to a reduction of the cross section for missing momenta below 0.3 $\gevc$ and a strong enhancement for $\ppm > 0.3$ $\gevc$. FSI contributions are maximal at a neutron angle between 65$\degree$
 and 75$\degree$ depending on the missing momentum (with increasing missing momentum the `peak' tends to shift toward smaller angles). This general behavior is well described by the eikonal based calculations.
For recoil angles above  75$\degree$ FSI seem to decrease and there appears a missing momenta region where the experimental cross sections are systematically above the calculations. The range of missing momenta where this occurs changes with recoil angle. At 75$\degree$ the region starts at $\ppm > 0.45$ $\gevc$ while for 95$\degree$ it starts at $\ppm > 0.3$ $\gevc$ and at 105$\degree$ it starts already at $\ppm > 0.2$ $\gevc$. At $\thetanq > 125\degree$ the effect gradually diminished but the cross sections remain systematically above the MS FSI CDB, MS FSI V18 and FJO~FSI~CDB calculations while the JML results show a smaller effect overall and start to over-predict the data for larger recoil angles.

\begin{figure*}[htp]
\includegraphics[width=0.95\textwidth,clip=true]{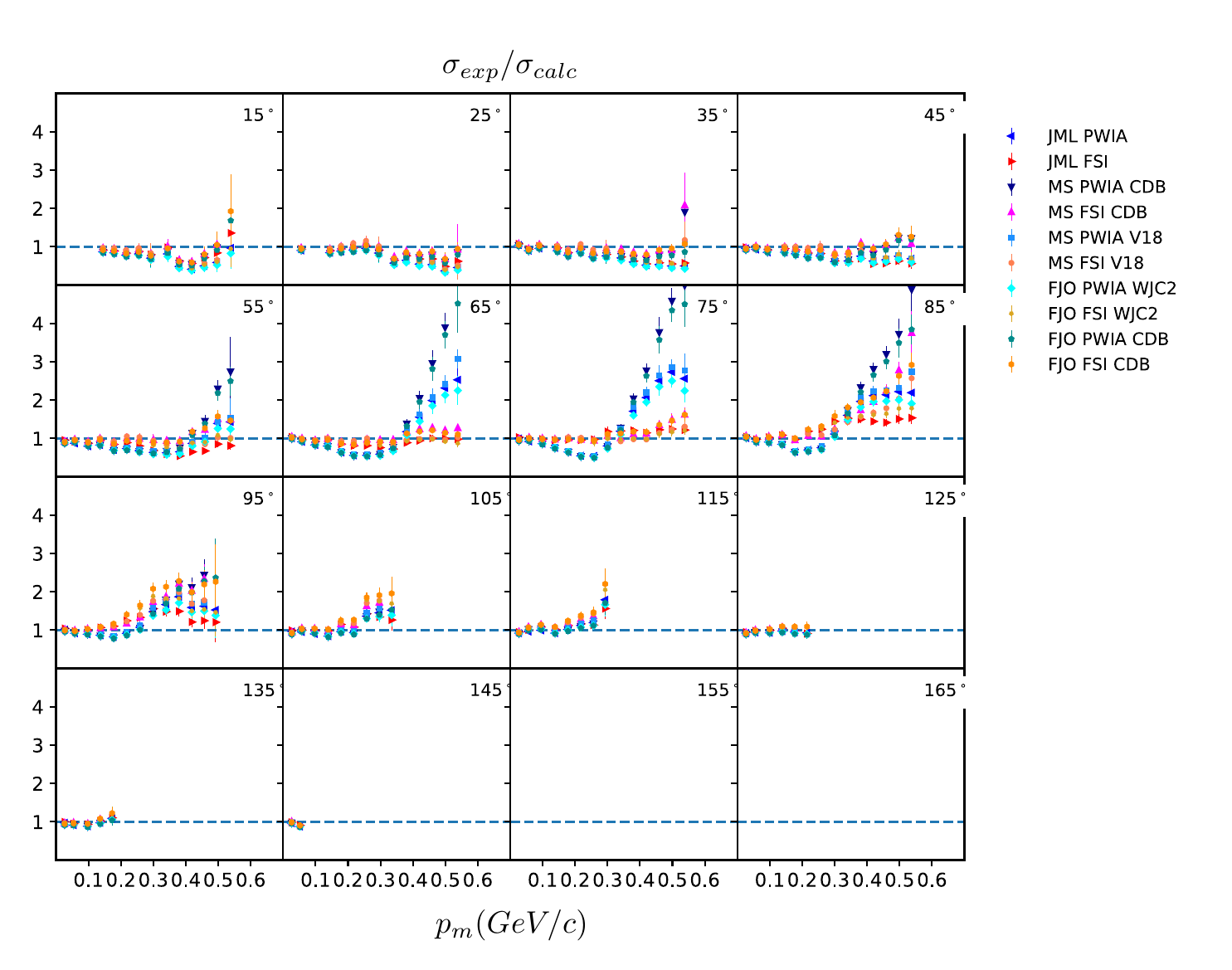}
\caption{ (Color online) Ratios of the experimental to calculated cross sections for a momentum transfer of $\Qtwo = 3.5$ $\gevctwo$. See Fig.~\ref{fig:pm_ratio_q1} for an explanation of the legends. The angles indicate the neutron recoil angle with respect to $\qvec$ for each data set.}
\label{fig:pm_ratio_q3}
\end{figure*}

The ratio for the highest momentum transfer data set ($\Qtwo = 3.5$ $\gevctwo$) is shown in Fig.~\ref{fig:pm_ratio_q3}. At this momentum transfer the eikonal approximation is expected to be valid over the entire kinematic range covered by this setting. Similar to the 2.1$\gevctwo$ data FSI are predicted to be small for recoil angles below 55$\degree$. All calculations agree within 10 - 20\% with the data for missing momenta $\ppm \leq 0.2$ $\gevc$.  For larger missing momenta in this small recoil angle range the MS FSI CDB and FJO FSI CDB calculations reproduce the data best compared to the other calculations which as in the $\Qtwo = 2.1$ $\gevctwo$ case over-predict the cross sections for missing momenta above 0.3$\gevc$.  It is also noteworthy that the MS FSI CDB and FJO FSI CDB calculations produce almost identical results for both the 2.1$\gevctwo$  and the 3.2.1$\gevctwo$ momentum transfer settings.

At recoil angles above $55\degree$ FSI start to dominate above missing momenta of 0.5 $\gevc$ similar to what has been observed at $\Qtwo = 2.1$ $\gevctwo$. FSI appear to contribute maximally over the full angular range for $\thetanq = 75\degree$. At this angle there is virtually no difference between the various calculations indicating that the cross sections are actually dominated by the low missing momentum part of the momentum distribution where the Paris, V18, WJC2 and CD-Bonn momentum distributions are very similar. At larger recoil angles experimental cross sections are again larger than predicted and show a similar behavior as has been observed at the lower momentum transfer setting hinting at significant contributions from intermediate isobar configurations for missing momenta above 0.25 $\gevc$. For angles larger than $95\degree$ the highest missing momentum measured was unfortunately limited by the momentum reach of the hadron spectrometer.

Data files containing the full and reduced cross sections  of the data shown in Figs.~\ref{fig:pm_dist}, ~\ref{fig:pm_ratio_q1},  ~\ref{fig:pm_ratio_q2} and ~\ref{fig:pm_ratio_q3} together with detailed kinematic variables for each data point can be found in the supplemental material~\cite{supp}.

\clearpage
%
\section{Summary and Conclusion}\label{L_summary}

We measured $\deep$ cross sections at invariant momentum transfers square of 0.8, 2.1 and 3.5 $\gevctwo$ 
covering a sufficient range of proton kinematics at each $\Qtwo$
that allowed us to investigate this reaction as a function of missing momenta as well as a function of the neutron recoil angle $\thetanq$ for each momentum transfer.
We experimentally confirmed that the generalized eikonal approximation, which predicts a strong angular dependence of FSI contributions, is not valid at$\gevctwo = 0.8$ $\gevctwo$ while it appears to be valid at 2.1 $\gevctwo$ for $\thetanq \geq 25\degree$ and $\ppm \leq 0.5$ $\gevc$ and at 3.5 $\gevctwo$ for the full kinematic range measured. 
We also found that for a range of (recoil) neutron angles between $25\degree$ and $45\degree$, FSI are minimal and virtually independent of missing momentum  providing a more direct experimental access to the pre-existing momentum distribution in the deuteron. We were therefore able to determine reduced cross sections (or experimental momentum distributions) at 2.1 and 3.5 $\gevctwo$  for a missing momentum range of $0.025 \leq \ppm \leq 0.5$ $\gevc$ where FSI contribute minimally. We found that for missing momenta below 0.3 $\gevc$ and neutron angles smaller that 45$\degree$ all calculations represent the data reasonable well (within $\sim$ 25\%). For missing momenta above 0.4 $\gevc$ and the same recoil angle range calculations based on CD-Bonn wave functions were systematically in better agreement with the data than calculations based on V18 or WJC2 wave functions. 

 
\begin{acknowledgments}

The authors acknowledge the outstanding efforts of the staff of the Accelerator
and Physics Divisions at Jefferson Lab who made this experiment
possible. The authors also thank Misak Sargsian for his many suggestions and for carefully reading the manuscript. This work was supported in part by the Department of Energy
under contracts DE-SC0013620, DE-FG02-99ER41065, DE-AC02-06CH11357, the Italian
Istituto Nazionale di Fisica Nucleare, the French Centre National de
la Recherche Scientifique and the National Science
Foundation. This material is based upon work supported by the U.S. Department of Energy, 
Office of Science, Office of Nuclear Physics under contract DE-AC05-06OR23177

\end{acknowledgments}

\onecolumngrid
\appendix

\section{Kinematic Settings}
\label{app:A}

\subsection{Overview and Labeling}

\begin{figure*}[htb]
\centering
\includegraphics[width=.85\textwidth,clip=true]{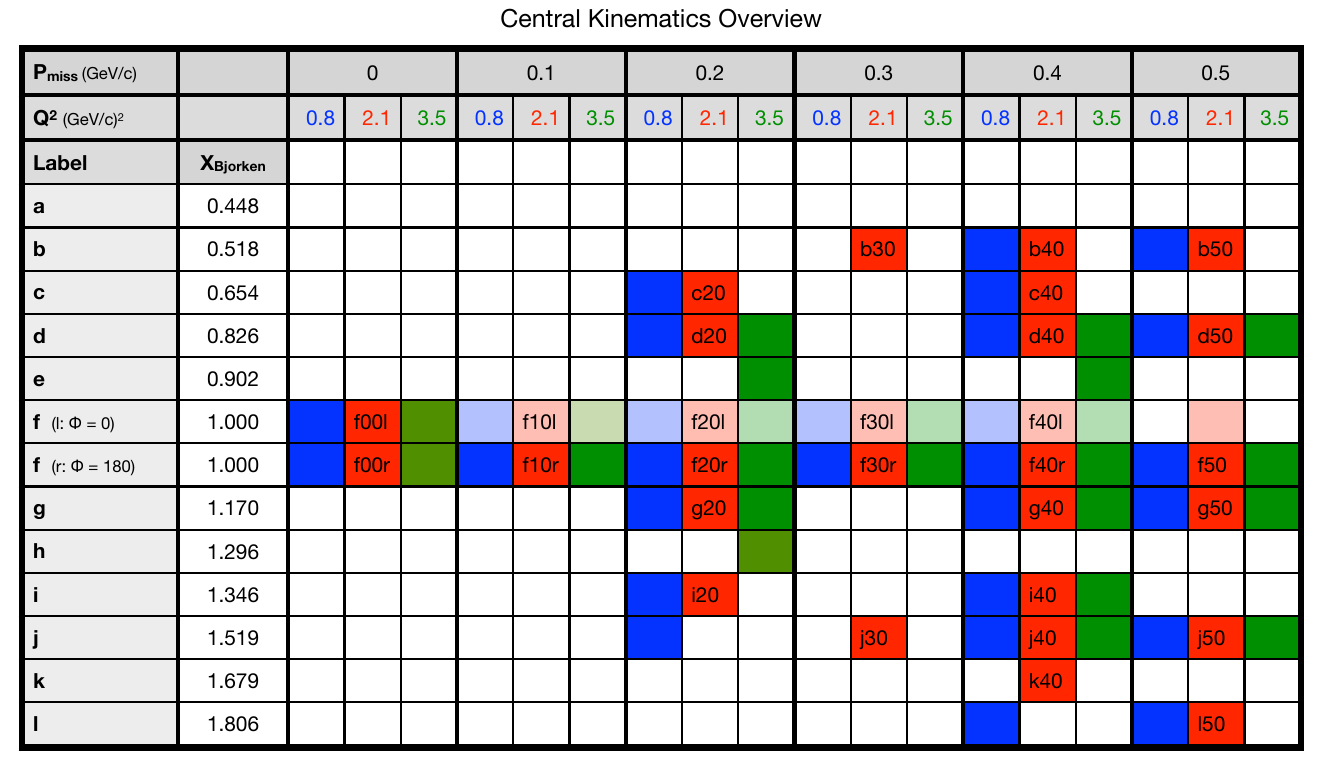}
\caption{ (Color online) Overview of kinematic spectrometer settings together with an example of labels used
for the $\Qtwo = 2.1$ settings.}
\label{fig:kin_overview}
\end{figure*}

The experiment described in this article is the result of the combination of 65 different
spectrometer settings. The numerical results presented in the supplemental material~\cite{supp} provide the experimental cross sections, the corresponding kinematics, and the central kinematic spectrometer setting for each data point calculated as described in the article. The central spectrometer setting used for each data point is provided as well.  An overview of the various central kinematic spectrometer settings is shown in Fig.~\ref{fig:kin_overview}. 
A kinematic spectrometer setting is defined by the central $\Qtwo$-value, the central Bjoken-x ($x_B$) value and the central missing momentum value. Settings with a common $\xbj$ values lie on a row while settings with common $\ppm$ values are on a column of the table shown in Fig.~\ref{fig:kin_overview}. The $\xbj$ values are further identified by the letters {\it{a}} through {\it{l}} as given by the columns "Label" and "${\rm x_{Bjorken}}$". 

The kinematic label is formed by combining the $\xbj$-letter with the missing momentum value in $\gevc\times 100$. White fields in the table indicate that no data have been taken at this setting. A solid colored cell indicates that this setting has been measured and been used in this work. The $\Qtwo$-row of column labels provides the  $\Qtwo$-values used for each $\ppm$ value and are color coded as follows: \textcolor{blue}{blue: 0.8 $\gevctwo$}, \textcolor{red}{red: 2.1 $\gevctwo$} and \textcolor{green}{green: 3.5 $\gevctwo$}. A light color  means data have been taken but not used here. 

As an example the setting for a momentum transfer $\Qtwo = 3.5$ $\gevctwo$, a missing momentum of 0.4 $\gevc$ and $\xbj = 1.59$ is labeled {\bf j40} and data have been taken at this setting, while no data have been taken for $\xbj = 1.346$ and $\ppm = 0.3$ $\gevc$.

\subsection{Central Spectrometer Settings}

The tables below list all the central spectrometer settings used. The kinematic setting used for the lowest momentum transfer $\Qtwo = 0.8$ $\gevctwo$ is shown in table~\ref{tab:kin_cen_q1}, the one for $\Qtwo = 2.1$ $\gevctwo$ in table~\ref{tab:kin_cen_q2} and the one for the highest momentum transfer $\Qtwo = 3.5$ $\gevctwo$ is shown in table~\ref{tab:kin_cen_q3}. The column labeled {\it Label} is the label for this kinematic setting and corresponds to the one in the left most column in Fig.~\ref{fig:kin_overview}. $\xbj$ refers to Bjoken-x, $E_i$ is the incident electron energy, $\theta_e$ the electron spectrometer central ray angle with respect to the beam, $E_f$ the  electron spectrometer central momentum, $\theta_p$ the proton spectrometer central ray angle with respect to the beam and $P_f$ is the proton spectrometer central momentum. Note that the central spectrometer settings should not be used to calculate reaction kinematics.  The kinematic data for  each kinematic bin and  provided in the numerical data and are described below.

\vspace{0.5 in}
%
\begin{table}[!h]
\centering 
\begin{tabular}{|c|c|c|c|c|c|c|} 
\hline 
{\it Label}& $\xbj$ &  $E_i$      & $\theta_e$ & $E_f$ & $\theta_p$ & $P_f$   \\ \hline 
                       &         & {\small (GeV)} & {\small $(\degree)$} & {\small (GeV)} & {\small $(\degree)$} & {\small $\rm (GeV/c) $}\\ \hline 
b40  & 0.517 & 2.844 & 21.51 & 2.020 & 53.54 & 1.390   \\ 
b50  & 0.517 & 2.844 & 21.51 & 2.020 & 59.54 & 1.338   \\ 
c20  & 0.641 & 2.844 & 20.22 & 2.202 & 51.92 & 1.239   \\ 
c40  & 0.641 & 2.844 & 20.22 & 2.202 & 65.00 & 1.162   \\ 
d20  & 0.823 & 2.843 & 20.03 & 2.325 & 61.09 & 1.081   \\ 
d40  & 0.825 & 2.843 & 20.03 & 2.326 & 73.17 & 0.998   \\ 
d50  & 0.823 & 2.843 & 20.03 & 2.325 & 79.42 & 0.939   \\ 
f00  & 0.993 & 2.843 & 19.65 & 2.414 & 54.92 & 0.993   \\ 
f10r  & 0.993 & 2.844 & 19.65 & 2.414 & 60.87 & 0.981   \\ 
f20r  & 0.994 & 2.843 & 19.65 & 2.414 & 66.71 & 0.960   \\ 
f30r  & 0.993 & 2.844 & 19.65 & 2.414 & 72.66 & 0.923   \\ 
f40r  & 0.993 & 2.843 & 19.65 & 2.414 & 78.80 & 0.873   \\ 
f50  & 0.994 & 2.843 & 19.65 & 2.414 & 85.23 & 0.808   \\ 
g20  & 1.164 & 2.843 & 19.40 & 2.477 & 69.47 & 0.870   \\ 
g40  & 1.164 & 2.843 & 19.40 & 2.477 & 82.00 & 0.778   \\ 
g50  & 1.164 & 2.843 & 19.40 & 2.477 & 88.53 & 0.708   \\ 
i20  & 1.341 & 2.843 & 19.21 & 2.525 & 69.80 & 0.797   \\ 
i40  & 1.340 & 2.843 & 19.21 & 2.525 & 83.38 & 0.700   \\ 
j20  & 1.514 & 2.843 & 19.08 & 2.561 & 65.40 & 0.740   \\ 
j40  & 1.509 & 2.843 & 19.05 & 2.561 & 83.05 & 0.638   \\ 
j50  & 1.509 & 2.843 & 19.05 & 2.561 & 89.29 & 0.557   \\ 
l40  & 1.803 & 2.843 & 18.91 & 2.607 & 77.83 & 0.553   \\ 
l50  & 1.803 & 2.843 & 18.91 & 2.607 & 82.67 & 0.462   \\ 
\hline 
\end{tabular} 
\caption{\label{tab:kin_cen_q1} Central Spectrometer Settings for $\Qtwo = 0.8\ \gevctwo$} 
\end{table} 

%
\begin{table} 
\centering 
\begin{tabular}{|c|c|c|c|c|c|c|} 
\hline 
{\it Label}& $\xbj$ &  $E_i$      & $\theta_e$ & $E_f$ & $\theta_p$ & $P_f$   \\ \hline 
                       &         & {\small (GeV)} & {\small $(\degree)$} & {\small (GeV)} & {\small $(\degree)$} & {\small $\rm (GeV/c) $}\\ \hline 
b30  & 0.519 & 4.701 & 24.18 & 2.545 & 24.04 & 2.899   \\ 
b40  & 0.519 & 4.701 & 24.18 & 2.545 & 30.00 & 2.862   \\ 
b50  & 0.519 & 4.701 & 24.18 & 2.545 & 33.21 & 2.817   \\ 
c20  & 0.667 & 4.703 & 22.15 & 3.026 & 31.42 & 2.416   \\ 
c40  & 0.667 & 4.703 & 22.15 & 3.026 & 40.48 & 2.351   \\ 
d20  & 0.826 & 4.703 & 21.04 & 3.348 & 42.50 & 2.067   \\ 
d40  & 0.826 & 4.703 & 21.04 & 3.348 & 48.86 & 2.000   \\ 
d50  & 0.826 & 4.703 & 21.04 & 3.348 & 51.91 & 1.952   \\ 
f00  & 0.999 & 4.703 & 20.34 & 3.582 & 42.72 & 1.836   \\ 
f10r  & 0.999 & 4.703 & 20.34 & 3.582 & 45.97 & 1.824   \\ 
f20r  & 0.999 & 4.703 & 20.34 & 3.582 & 49.10 & 1.806   \\ 
f30r  & 0.999 & 4.703 & 20.34 & 3.582 & 52.23 & 1.777   \\ 
f40r  & 0.999 & 4.703 & 20.34 & 3.582 & 55.36 & 1.738   \\ 
f50r  & 0.999 & 4.703 & 20.34 & 3.582 & 58.51 & 1.689   \\ 
g20  & 1.169 & 4.703 & 19.88 & 3.746 & 52.77 & 1.619   \\ 
g40  & 1.169 & 4.703 & 19.88 & 3.746 & 59.61 & 1.549   \\ 
g50  & 1.170 & 4.703 & 19.88 & 3.747 & 62.90 & 1.498   \\ 
i20  & 1.348 & 4.703 & 19.55 & 3.873 & 51.21 & 1.498   \\ 
i40  & 1.348 & 4.703 & 19.55 & 3.873 & 61.96 & 1.398   \\ 
j30  & 1.520 & 4.703 & 19.31 & 3.967 & 54.24 & 1.326   \\ 
j40  & 1.520 & 4.703 & 19.31 & 3.967 & 62.09 & 1.283   \\ 
j50  & 1.520 & 4.703 & 19.31 & 3.967 & 66.27 & 1.229   \\ 
k40  & 1.679 & 4.703 & 19.15 & 4.036 & 56.81 & 1.196   \\ 
l50  & 1.812 & 4.703 & 19.03 & 4.086 & 58.42 & 1.076   \\ 
\hline 
\end{tabular} 
\caption{\label{tab:kin_cen_q2} Central Spectrometer Settings for $\Qtwo = 2.1\ \gevctwo$} 
\end{table} 

\vspace{0.5 in}
\begin{table} 
\centering 
\begin{tabular}{|c|c|c|c|c|c|c|} 
\hline 
{\it Label}& $\xbj$ &  $E_i$      & $\theta_e$ & $E_f$ & $\theta_p$ & $P_f$   \\ \hline 
                       &         & {\small (GeV)} & {\small $(\degree)$} & {\small (GeV)} & {\small $(\degree)$} & {\small $\rm (GeV/c) $}\\ \hline 
d20  & 0.828 & 5.009 & 29.23 & 2.751 & 30.59 & 3.031   \\ 
d40  & 0.828 & 5.009 & 29.23 & 2.751 & 34.97 & 2.968   \\ 
d50  & 0.828 & 5.009 & 29.23 & 2.751 & 37.02 & 2.923   \\ 
e20  & 0.902 & 5.008 & 28.30 & 2.934 & 33.78 & 2.839   \\ 
e40  & 0.902 & 5.009 & 28.30 & 2.934 & 38.02 & 2.776   \\ 
f00b  & 1.002 & 5.008 & 27.33 & 3.141 & 32.90 & 2.650   \\ 
f10r  & 1.002 & 5.008 & 27.33 & 3.141 & 35.16 & 2.637   \\ 
f20r  & 1.002 & 5.008 & 27.33 & 3.141 & 37.32 & 2.620   \\ 
f30r  & 1.002 & 5.008 & 27.33 & 3.141 & 39.46 & 2.593   \\ 
f40r  & 1.002 & 5.008 & 27.33 & 3.141 & 41.59 & 2.556   \\ 
f50  & 1.002 & 5.008 & 27.33 & 3.141 & 43.72 & 2.510   \\ 
g20  & 1.175 & 5.008 & 26.20 & 3.415 & 41.36 & 2.327   \\ 
g40  & 1.175 & 5.008 & 26.20 & 3.415 & 46.21 & 2.262   \\ 
g50  & 1.175 & 5.008 & 26.20 & 3.415 & 48.48 & 2.215   \\ 
h20  & 1.296 & 5.009 & 25.63 & 3.564 & 40.93 & 2.166   \\ 
i40  & 1.355 & 5.008 & 25.42 & 3.626 & 48.95 & 2.031   \\ 
j40  & 1.529 & 5.008 & 24.88 & 3.783 & 48.61 & 1.856   \\ 
j50  & 1.529 & 5.008 & 24.88 & 3.783 & 52.50 & 1.808   \\ 
\hline 
\end{tabular} 
\caption{\label{tab:kin_cen_q3} Central Spectrometer Settings for $\Qtwo = 3.5\ \gevctwo$} 
\end{table}

\section{Numerical Data}
\label{app:B}

\begin{figure*}[htb]
\centering
\includegraphics[width=.85\textwidth,clip=true]{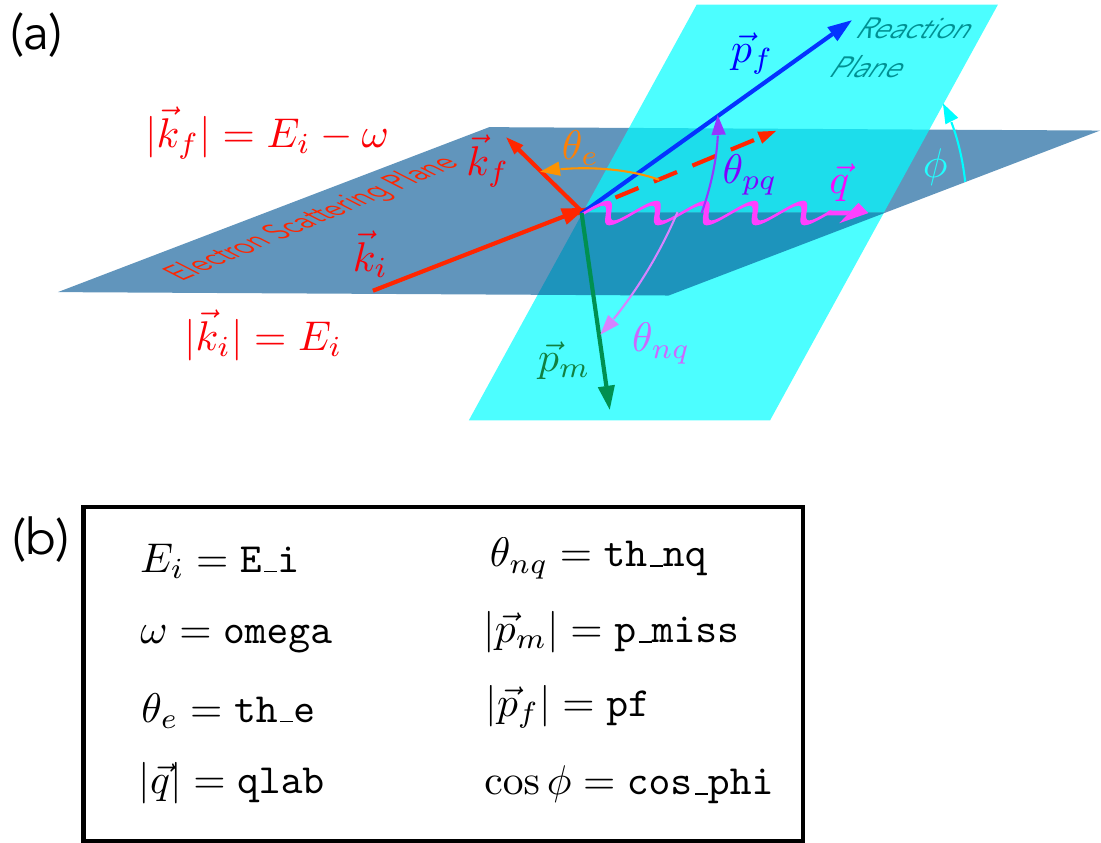}
\caption{ (Color online) (a) Kinematic variables of the $\deep$ reaction. (b) Correspondence between the variable names provided in the data files and the kinematic variables defined in (a).}
\label{fig:deep_reaction}
\end{figure*}

The  numerical results are contained in the compressed zip file \texttt{numerical\_results.zip}. Extracting this file creates a directory called \texttt{numerical\_results} containing the cross section and kinematics data for all data presented in the article. The data files are sorted according to their $\Qtwo$ values using the following naming convention: the labels \texttt{q1}, \texttt{q2} and \texttt{q3} refer to the $\Qtwo$ values of 0.8 $\gevctwo$ for \texttt{q1}, 2.1 $\gevctwo$ for \texttt{q2} and 3.5 $\gevctwo$ for \texttt{q3}. The kinematic variables provided and their relation to the $\deep$ reaction are illustrated in figure~\ref{fig:deep_reaction}. In panel (a) the red arrows labelled $\vec{k}_i$ and $\vec{k}_f$ correspond the to the incident and scattered electron respectively. The dashed red arrow represents electron beam direction. The wavy pink line, labelled $\vec{q}$, represents the 3-momentum transfer. 
The blue arrow ($\vec{p}_f$) represents the ejected proton and the green one ($\vec{p}_m$) the recoiling neutron or the missing momentum. Panel (b) lists the kinematic variables provided in the data files and their correspondence to the variables in panel (a). The names in the data files are written in \texttt{typewriter font} . $\theta_e$ is the electron scattering angle with respect to the beam,  $\thetapq$ is the angle of the ejected proton with respect to the momentum, transfer $\qvec$, $\phi$ is the angle between the electron scattering plane (defined by $\vec{k}_i$ and $\vec{k}_f$)  and the proton reaction plane (defined by $\vec{p}_f$ and $\qvec$) and $\thetanq$ is the angle of the recoiling neutron with respect to $\qvec$.   

\subsection{Angular Distributions}
The numerical data for the angular distributions are given for the missing momentum bins centered around 0.1, 0.2, 0.4 and 0.5 $\gevc$. The corresponding data files are named as follows: \texttt{ang\_dist\_qX\_pm\_0.XX.data} where \texttt{qX} refers to the $\Qtwo$ label (\texttt{q1}, \texttt{q2}, \texttt{q3}) defined above, and \texttt{0.XX} refers to the missing momentum value in $\gevc$.

As an example the file in \texttt{ang\_dist\_q2\_pm\_0.40.data} contains the angular distribution data for the momentum transfer $\Qtwo = 2.1$ $\gevctwo$ and $\ppm = 0.4$ $\gevctwo$. 

The data files are ASCII files containing multiple columns of data as well as a header section describing the units used and the meaning of each data column. The header information in the data files are those lines starting with a hashtag (\texttt{\#}) similar to comment lines in UNIX shell scripts. 

Each data line below the last header line provides the experimental cross section and PWIA ratio for a specific $\thetanq$ angle bin. It further provides the full kinematic setting for this data point as well as the bin-centering correction factor. Often multiple kinematic settings contribute to the same  $\thetanq$ angle bin. Each contributing kinematic setting is listed individually but they can be identified by having the same data number (\texttt{Nr}) and the same average $\thetanq$ value (\texttt{th\_nq\_av}). The column named \texttt{KinId} contains the kinematic label associated with a data row.  Please consult the header information for details on the data columns. All angles are in the Lab-frame.

\subsection{Missing Momentum Distributions}
The numerical data for the missing momentum distributions are given for fixed recoil angles $\thetanq$ and fixed $\Qtwo$ value.  The momentum distribution result files have the naming convention \newline \texttt{pm\_distribution\_results\_qX\_XXX.00.data} where where \texttt{qX} again refers to the $\Qtwo$ label and the \texttt{XXX.00} part refers to the $\thetanq$-angle for this data set. 

As an example the file \texttt{pm\_distribution\_results\_q3\_85.00.data} contains the numerical results for the missing momentum distribution measured at $\Qtwo = 3.5$ $\gevctwo$ at an angle $\thetanq = 85.0\degree$. As for the angular distributions the ASCII files contain the information explaining the meaning of each data column.

The format of the data file is similar to the one for angular distributions. The kinematic setting contributing to each row is in the column named \texttt{bin\_id}.

%
\bibliography{all_1-1}

@preamble{"\hyphenation{Post-Script Sprin-ger}"}

@string{jpg = {J. Phys. G}}

@string{nima = {Nucl. Instr. and Meth. A}}

@string{npa = {Nucl. Phys. A}}

@string{plb = {Phys. Lett. B}}

@string{prc = {Phys. Rev. C}}

@string{prd = {Phys. Rev. D}}

@string{prl = {Phys. Rev. Lett.}}

@misc{supp,
	note = {See Supplemental Material at [URL will be inserted by publisher] for the file numerical\_results.zip containing numerical data.}}

@article{Arrington:2011xs,
	archiveprefix = {arXiv},
	author = {Arrington, J. and Higinbotham, D. W. and Rosner, G. and Sargsian, M.},
	doi = {10.1016/j.ppnp.2012.04.002},
	eprint = {1104.1196},
	journal = {Prog. Part. Nucl. Phys.},
	pages = {898--938},
	primaryclass = {nucl-ex},
	reportnumber = {PHY-12946-ME-2011, JLAB-PHY-11-1329},
	title = {{Hard probes of short-range nucleon-nucleon correlations}},
	volume = {67},
	year = {2012}}

@article{Fomin:2011ng,
	archiveprefix = {arXiv},
	author = {Fomin, N. and others},
	doi = {10.1103/PhysRevLett.108.092502},
	eprint = {1107.3583},
	journal = {Phys. Rev. Lett.},
	pages = {092502},
	primaryclass = {nucl-ex},
	reportnumber = {JLAB-PHY-11-1408},
	title = {{New measurements of high-momentum nucleons and short-range structures in nuclei}},
	volume = {108},
	year = {2012}}

@article{Boeglin:2015cha,
	archiveprefix = {arXiv},
	author = {Boeglin, Werner and Sargsian, Misak},
	doi = {10.1142/S0218301315300039},
	eprint = {1501.05377},
	journal = {Int. J. Mod. Phys. E},
	number = {03},
	pages = {1530003},
	primaryclass = {nucl-ex},
	reportnumber = {FIU-NUPAR-01-2015-1},
	title = {{Modern Studies of the Deuteron: from the Lab Frame to the Light Front}},
	volume = {24},
	year = {2015}}

@article{JML_CJP_1984_1046,
	author = {Laget, J. M.},
	doi = {10.1139/p84-142},
	eprint = {https://doi.org/10.1139/p84-142},
	journal = {Canadian Journal of Physics},
	number = {11},
	pages = {1046-1063},
	title = {Photo- and electrodisintegration of few-body systems at intermediate energies: an introduction},
	url = {https://doi.org/10.1139/p84-142},
	volume = {62},
	year = {1984}}

@article{JML_NPA_1978_265,
	author = {J.M. Laget},
	doi = {https://doi.org/10.1016/0375-9474(78)90590-0},
	issn = {0375-9474},
	journal = {Nuclear Physics A},
	number = {3},
	pages = {265-290},
	title = {Electromagnetic properties of the πNN system (III). The γD → pn reaction},
	url = {https://www.sciencedirect.com/science/article/pii/0375947478905900},
	volume = {312},
	year = {1978}}

@article{GROSS2007176,
	author = {Franz Gross and Alfred Stadler},
	doi = {https://doi.org/10.1016/j.physletb.2007.10.028},
	issn = {0370-2693},
	journal = {Physics Letters B},
	number = {4},
	pages = {176-179},
	title = {High-precision covariant one-boson-exchange potentials for np scattering below 350 MeV},
	url = {https://www.sciencedirect.com/science/article/pii/S0370269307012622},
	volume = {657},
	year = {2007}}

@article{PhysRevC.90.064006,
	author = {Ford, William P. and Jeschonnek, Sabine and Van Orden, J. W.},
	doi = {10.1103/PhysRevC.90.064006},
	issue = {6},
	journal = {Phys. Rev. C},
	month = {Dec},
	numpages = {11},
	pages = {064006},
	publisher = {American Physical Society},
	title = {Momentum distributions for $^{2}\mathrm{H}(e,{e}^{\ensuremath{'}}p)$},
	url = {https://link.aps.org/doi/10.1103/PhysRevC.90.064006},
	volume = {90},
	year = {2014}}

@article{CD_Bonn_Potential,
	author = {Machleidt, R.},
	doi = {10.1103/PhysRevC.63.024001},
	issue = {2},
	journal = {Phys. Rev. C},
	month = {Jan},
	numpages = {32},
	pages = {024001},
	publisher = {American Physical Society},
	title = {High-precision, charge-dependent Bonn nucleon-nucleon potential},
	url = {https://link.aps.org/doi/10.1103/PhysRevC.63.024001},
	volume = {63},
	year = {2001}}

@article{AV18_Potential,
	author = {Wiringa, R. B. and Stoks, V. G. J. and Schiavilla, R.},
	doi = {10.1103/PhysRevC.51.38},
	issue = {1},
	journal = {Phys. Rev. C},
	month = {Jan},
	numpages = {0},
	pages = {38--51},
	publisher = {American Physical Society},
	title = {Accurate nucleon-nucleon potential with charge-independence breaking},
	url = {https://link.aps.org/doi/10.1103/PhysRevC.51.38},
	volume = {51},
	year = {1995}}

@article{Paris_Potential,
	author = {Lacombe, M. and Loiseau, B. and Richard, J. M. and Mau, R. Vinh and C\^ot\'e, J. and Pir\`es, P. and de Tourreil, R.},
	doi = {10.1103/PhysRevC.21.861},
	issue = {3},
	journal = {Phys. Rev. C},
	month = {Mar},
	numpages = {0},
	pages = {861--873},
	publisher = {American Physical Society},
	title = {Parametrization of the Paris $N\ensuremath{-}N$ potential},
	url = {https://link.aps.org/doi/10.1103/PhysRevC.21.861},
	volume = {21},
	year = {1980}}

@article{Sabine1,
	author = {A.~Bianconi and S.~Jeschonnek and N.~N.~Nikolaev and B.~G.~Zakharov},
	journal = PLB,
	pages = 13,
	volume = 343,
	year = 1995}

@article{noredepn,
	author = {M.~M.~Sargsian},
	journal = PRC,
	pages = {014612},
	volume = 82,
	year = 2010}

@article{JW10,
	author = {S.~Jeschonnek and J.~W.~{Van Orden}},
	journal = PRC,
	pages = {014008},
	volume = 81,
	year = 2010}

@article{Arrington_1998ps,
	author = {Arrington, J. and others},
	journal = {Phys. Rev. Lett.},
	pages = {2056-2059},
	title = {{Inclusive electron nucleus scattering at large momentum transfer}},
	volume = {82},
	year = {1999}}

@article{MGWvO,
	author = {M.~Gar\c{c}on and J.~W.~{Van Orden}},
	journal = {Adv.~Nucl.~Phys.},
	pages = 293,
	volume = 26,
	year = 2002}

@article{Rock92,
	author = {{S. Rock} and others},
	journal = PRD,
	pages = {24},
	volume = 46,
	year = 1992}

@article{Rock82,
	author = {P.~E.~Bosted and R.~G.~Arnold and S.~Rock and Z.~M.~Szalata},
	journal = PRL,
	pages = 1380,
	volume = 49,
	year = 1982}

@article{edsacley,
	author = {{A.~Bussiere} and others},
	journal = NPA,
	pages = {349},
	volume = 365,
	year = 1981}

@article{Alexa,
	author = {{L. C. Alexa} and others},
	journal = PRL,
	pages = {1374},
	volume = 82,
	year = 1999}

@article{halla,
	author = {{John Alcorn} and others},
	journal = {NIMA},
	pages = {294},
	title = {Basic Instrumentation for Hall A at Jefferson Jab},
	volume = 522,
	year = 2004}

@article{Boe08,
	author = {Boeglin, W.U. and others},
	journal = PRC,
	pages = {054001},
	volume = {78},
	year = {2008}}

@article{bl98,
	author = {{K.I.~Blomqvist} and others},
	journal = PLB,
	pages = 33,
	volume = 424,
	year = 1998}

@article{Ulm02,
	author = {{P.E~Ulmer} and others},
	journal = PRL,
	pages = {062301-1},
	volume = 89,
	year = 2002}

@article{FMGSS95,
	author = {{L.L.~Frankfurt} and {W.G.~Greenberg} and {J.A.~Miller} and {M.M.~Sargsian} and {M.I.~Strikman}},
	journal = {Z. Phys. A},
	pages = 97,
	volume = 352,
	year = 1995}

@article{FSS96,
	author = {{L.L.~Frankfurt} and {M.M.~Sargsian} and {M.I.~Strikman}},
	journal = PRC,
	pages = 1124,
	volume = 56,
	year = 1997}

@article{La05,
	author = {{J.M.~Laget}},
	journal = PLB,
	pages = {49},
	volume = {609},
	year = 2005}

@article{deFor83,
	author = {{T.~de~Forest}},
	journal = NPA,
	pages = 232,
	volume = 392,
	year = 1983}

@article{Kap05,
	author = {{C.~Ciofi~degli~Atti} and {L.~P.~Kaptari}},
	journal = PRC,
	pages = 024005,
	volume = 71,
	year = 2005}

@article{JW08,
	author = {S. Jeschonnek and J.~W.~{Van~Orden}},
	journal = PRC,
	pages = 014007,
	volume = 78,
	year = 2008}

@article{JW09,
	author = {S.~Jeschonnek and J.~W.~{Van Orden}},
	journal = PRC,
	pages = 054001,
	volume = 80,
	year = 2009}

@article{Ciofi08,
	author = {{C.~Ciofi~degli~Atti} and L.~P.~Kaptari},
	journal = PRL,
	pages = 122301,
	volume = 100,
	year = 2008}

@article{Kim3,
	author = {K.~S.~Egiyan and others},
	journal = PRL,
	pages = 262502,
	volume = 98,
	year = 2007}

@article{treeview,
	author = {M.~M.~Sargsian},
	journal = {Int. J. Mod. Phys. E},
	pages = 405,
	volume = 10,
	year = 2001}

@article{Jesch2001,
	author = {S.~Jeschonnek},
	journal = PRC,
	pages = 034609,
	volume = 63,
	year = 2001}

@article{hnm,
	author = {M.~M.~Sargsian and others},
	journal = JPG,
	pages = {R1},
	volume = 29,
	year = 2003}

@article{Arr04,
	author = {J.~Arrington},
	journal = PRC,
	pages = {022201(R)},
	volume = 69,
	year = 2004}

@article{Rfunc,
	author = {V. L. Rvachev and T. I. Sheiko},
	journal = {Appl. Mech. Rev.},
	pages = 151,
	volume = 48,
	year = 1995}

@article{SIMC,
	author = {https://hallcweb.jlab.org/wiki/index.php/Monte\_Carlo},
	title = {Computer Program SIMC},
	year = {2009}}

@article{MCEEP,
	author = {http://hallaweb.jlab.org/software/mceep/mceep.html},
	title = {Computer Program MCEEP},
	year = {2006}}

@article{Rva05,
	author = {M.~Rvachev and F.~Benmokhtar and E.~Penel-Nottaris and others},
	journal = PRL,
	pages = {192302},
	title = {The quasielastic $^3$He(e,e$'$p)d reaction at Q$^2$~=~1.5~GeV$^2$ for recoil momenta up to 1~GeV/$c$},
	volume = {94},
	year = {2005}}

@article{GilGro02,
	author = {R.~Gilman and F.~Gross},
	journal = JPG,
	pages = {R37},
	volume = {28},
	year = {2002}}

@article{Ent01,
	author = {Ent, R. and Filippone, B. W. and Makins, N. C. R. and Milner, R. G. and O'Neill, T. G. and Wasson, D. A.},
	doi = {10.1103/PhysRevC.64.054610},
	journal = PRC,
	month = {Oct},
	number = {5},
	numpages = {23},
	pages = {054610},
	publisher = {American Physical Society},
	title = {Radiative corrections for $(e,e^{'{}}p)$ reactions at GeV energies},
	volume = {64},
	year = {2001}}
\end{document}